\documentclass[a4paper]{amsart}

\usepackage[english]{babel}
\usepackage[T1]{fontenc}
\usepackage{amsmath,amsthm}
\usepackage{amsfonts}
\usepackage{graphicx}
\graphicspath{{./Figures/}}
\usepackage{xcolor}
\usepackage {subfigure}
\usepackage{setspace}
\usepackage[scr=boondoxo,scrscaled=1.05]{mathalfa}

\usepackage{enumitem}
\calclayout

\usepackage[colorinlistoftodos,prependcaption]{todonotes}

\newcommand{\revised}[1]{\textcolor{black}{#1}}
\newcommand{\revisedfurther}[1]{\textcolor{black}{#1}}

\newtheorem{theorem}{Theorem}

\newcommand{\be}{\begin{equation}}
\newcommand{\ee}{\end{equation}}

\def\var{\gamma}
\def\r{\rho}
\def\l{\lambda}
\def\t{\tau}

\def\a{\alpha}
\def\b{\beta}
\newcommand{\deRi}{\frac{\partial}{\partial R_i}}
\newcommand{\deRiRi}{\frac{\partial^2}{\partial R_i^2}}
\newcommand{\deri}{\frac{\partial}{\partial \r_i}}
\newcommand{\deriri}{\frac{\partial^2}{\partial \r_i^2}}
\newcommand{\deR}{\frac{\partial}{\partial R}}
\newcommand{\deRR}{\frac{\partial^2}{\partial R^2}}
\newcommand{\der}{\frac{\partial}{\partial r}}
\newcommand{\dr}{\frac{\partial}{\partial \r}}
\newcommand{\drr}{\frac{\partial^2}{\partial \r^2}}
\newcommand{\dt}{\frac{\partial}{\partial t}}

\def\R{\mathbb{R}}

\newcommand{\inr}{\int_\R}

\newcommand{\s}{{\sigma^2}}
\newcommand{\fe}{f_\var}

\newcommand{\Ri}{{R_i}}
\newcommand{\Rj}{{R_j}}
\newcommand{\ri}{{\r_i}}
\newcommand{\rj}{{\r_j}}
\newcommand{\Rit}{{R_i^*}}
\newcommand{\Rjt}{{R_j^*}}
\newcommand{\rit}{{\r_i^*}}
\newcommand{\rjt}{{\r_j^*}}

\newcommand{\mri}{\mathscr{m}_\ri(t)}
\newcommand{\mRi}{\mathscr{m}_\Ri(t)}
\newcommand{\ct}{\tilde{c}}
\newcommand{\D}{{\Omega}}
\newcommand{\HD}{H^1(\D)}
\newcommand{\HDd}{H^{-1}(\D)}
\newcommand{\LD}{L^2(\D)}
\newcommand{\LDt}[1]{L^2(0,T; #1)}
\newcommand{\HDt}[1]{H^1(0,T; #1)}
\newcommand{\K}{K_M[\gmu]}
\newcommand{\A}{L_M[\gmu]}

\newcommand{\sm}{\frac{\sigma^2}{2}}
\newcommand{\norm}[2]{\| #1 \|_{#2}}

\newcommand{\gt}{\tilde{g}}
\newcommand{\gmu}{g_{\mu}}
\newcommand{\gm}{g_*}

\title[Boltzmann and FP equations modelling the ELO rating system]{\revised{Boltzmann and Fokker-Planck equations modelling the Elo rating system with learning effects}}
\author{Bertram D\"uring}
\address[B. D\"uring]{Department of Mathematics, University of Sussex, Brighton, BN1 9QH, UK}
\email{bd80@sussex.ac.uk}
\author{Marco Torregrossa}
\address[M. Torregrossa]{Department of Mathematics, University of Pavia, via Ferrata 1, Pavia, 27100, Italy}
\email{marcotorr1986@gmail.com}

\author{Marie-Therese Wolfram}
\address[M.-T. Wolfram]{Mathematics Institute, University of Warwick, Coventry, CV4 7AL, UK and Radon Institute of Computational and Applied Mathematics, Altenbergerstr. 69, 4040 Linz, Austria}
\email{m.wolfram@warwick.ac.uk}

\begin{document}

\begin{abstract}
In this paper we propose and study a new kinetic rating model for a
large number of players, which is motivated by the well-known Elo
rating system. 
Each player is characterised by an
intrinsic strength and a rating, which are both updated after each game. We state and analyse the respective Boltzmann type equation and derive the corresponding
nonlinear, nonlocal Fokker-Planck equation. We investigate the existence of solutions to the Fokker-Planck equation and discuss their behaviour in the long time limit.  Furthermore,
we illustrate the dynamics of the Boltzmann and Fokker-Planck equation with various numerical experiments.
\end{abstract}

\maketitle

\section{Introduction}\label{s:introduction}

\noindent In 1950 the Hungarian physicist Arpad Elo developed a rating system
to calculate the relative skill level of players in competitor versus
competitor games, see \cite{ELO}. The Elo rating system was initially
used in chess competitions, but was quickly adopted by the US Chess
Federation as well as the World Chess Federation, and the
National Football Foundation. In June 2018, FIFA announced switching
their world football ranking to an Elo system, following two years of
reviews and studies of different alternatives. The Elo rating system assigns each player a rating, which is updated
according to the wins and losses as well as the difference of the ratings. It is hoped that the rating
converges to the relative strength level and is a valid measure of the player's skills. However, assigning an initial rating to a new player is a delicate issue, since it is not clear
 how an inaccurate initial rating influences the latter performance. Elo himself tried to validate the model using computational experiments, while Glickman used statistical techniques
to understand the dynamics \cite{Glickman}.
The first rigorous proof
of convergence of the ratings to the individual strength was presented by Junca and Jabin in \cite{JJ2015}, who 
introduced a continuous version of the Elo rating system. In this continuous model every player is
characterised by its intrinsic strength $\rho$ and rating
$R$. The intrinsic strength is fixed in time. If two
players with rating $R_i$ and $R_j$ meet in a game, their ratings
after the game,  $R_i^{*}$ and $R_j^{*}$ are given by
\begin{subequations}
\label{e:JJ}
\begin{align}
R_i^{*} &= R_i + K (S_{ij} - b(R_i-R_j)),\\
R_j^{*} &= R_j + K (-S_{ij} - b(R_j-R_i)).
\end{align}
\end{subequations}
In \eqref{e:JJ} the random variable $S_{ij}$ is the score result of the game, it takes
the value $1$ if player $i$ wins and the value $-1$ if player $j$
wins. The mean score (i.e.\ expected value of $S_{ij}$) is assumed to
be equal to $b(\rho_i-\rho_j)$, hence the result of each game depends on the difference of the player's intrinsic strengths.
The rating of each player in- or decreases proportionally with the
outcome of the game, relative to the predicted mean score $b(R_i-R_j)$. The speed of the adjustment is controlled by the
constant parameter $K$. The function $b$ is chosen in such a
way that extreme differences are moderated; a typical choice is
\begin{align}\label{function b}
b(z) &=\tanh(c z),
\end{align}
where $c$ is a suitably chosen positive constant. This choice weighs the impact
of the outcome with respect to the relative rating. If a player with a
high rating wins a game against a player with a low rating, the
players' ratings change little. However, if the player with the low
rating wins against a highly rated player, the ratings are
strongly adjusted. 

Junca and Jabin proposed the following Boltzmann type equation to describe the evolution of the distribution of players $f = f(r,t)$ with respect to their ratings
\begin{align}
\label{e:originalboltzmann}
\partial_t f(r,t) + \partial_r (a(f) f) = 0 \text{ with } a(f) = \int_{\mathbb{R}^2} w(r-r')(b(\rho-\rho') - b(r-r'))f(t,r',\rho') d\rho'dr'.
\end{align}
This equation describes a more general setup than in the microscopic
equations. Here two players only interact according to the interaction
rate function $w$,
which depends on the difference of their ratings. The function $w$ is
assumed to be even and nonnegative.
Junca and Jabin analysed the long time behaviour of solutions to 
\eqref{e:originalboltzmann}. They proved that in the case $w=1$, a so-called `all-play-all' tournament, the ratings converge
exponentially fast to the intrinsic strength. In the case of local interactions, that is individuals only play if their
ratings are close, the ratings may not converge to the intrinsic strength and the rating fails to give a fair representation
of the player's strength distribution.

Rather recently Krupp \cite{K2016} proposed an extension of the model by Jabin and
Junca \cite{JJ2015}. In her model not only the rating, but also the intrinsic strength changes as players continuously compete in games. 
In particular, she assumes that the intrinsic strength $\rho$ changes in every game according to
\begin{subequations}
\begin{align}
&\rho_i^{*} = \rho_i  + Z_{ij} \tilde{K} ,\\
&\rho_j^{*} = \rho_j + Z_{ij} \tilde{K},
\end{align}
\end{subequations}
where $\tilde{K}$ is a positive constant and $Z_{ij}$ takes the value $z_1 \in \mathbf{N}$ \revised{or} $z_2\in \mathbf{N}$. \revised{In case of a win the inner strength $\rho_i$ increases by $z_1 \tilde{K}$, in case of a loss by $z_2 \tilde{K}$. Hence if} $z_1 < z_2$ the looser benefits more from the game, while if $z_1 > z_2$ the winner learns more. If $z_1 = z_2$ both learn the same.
The corresponding Boltzmann type equation for the distribution of the players $f = f(r, \rho, t)$ with respect to their strength and rating reads as	
\begin{align}\label{e:boltzmannkrupp}
\partial_t f(r,\rho,t) + \partial_r(a(f) f) + \partial_\rho(c(f) f) = 0,
\end{align}
where 
\begin{align*}
a(f) = \int_{\mathbb{R}^2} w (r-r')[b(\rho-\rho')-b(r-r')] f(r',\rho',t) d\rho' dr'
\end{align*}
and 
\begin{align*}
c(f) = \int_{\mathbb{R}^2} w(r-r') [\frac{z_1}{2}(b(\rho-\rho') + 1) - \frac{z_2}{2}(b(r-r')-1)]f(r',\rho',t) d\rho' dr'.
\end{align*}
Krupp analysed the qualitative behaviour of solutions to \eqref{e:boltzmannkrupp}. 
Due to the continuous increase in strength, the ratings increase in time. Therefore, an appropriately shifted problem was studied, 
in which the ratings converged exponentially fast to the intrinsic strength in the case $w=1$.

In this paper we propose a more general approach to describe how a player's strength changes in encounters. We assume that 
individuals benefit from every game and increase their strength because of these interactions. 
However, the extent of the benefit depends on several factors -- first, players with a lower rating benefit more.
Second, the stronger the opponent, the more a win pushes the intrinsic strength. 
Furthermore, the individual performance changes due to small fluctuations, accounting for variations in the mental strength or personal 
fitness on a day. Based on the microscopic interaction laws we derive
the corresponding kinetic Boltzmann type and limiting Fokker-Planck equations and analyse their behaviour. 
In the case of no diffusion we can show that the strength and ratings of the appropriately shifted PDE converge, while we observe the 
formation of non-measure valued steady states in the case of diffusion. We illustrate our analytic results with numerical simulations
of the kinetic as well as the limiting Fokker-Planck equation. The simulations give important insights into the dynamics, especially
in situations where we are not able to prove rigorous results.  
The proposed interaction laws are a first step to develop and analyse more complicated rating models with dynamic strength. The next
developments of the model should include losses in the player's strength to ensure that the strength stays within certain bounds. \\

The kinetic description of the Elo rating system \revised{allowed Junca $\&$ Jabin to analyse } the qualitative behaviour of solutions.
In the last decades kinetic models have been used successfully to describe the behaviour of large multiagent systems in socio-economic 
applications. In all these applications interactions among individuals are modeled as `collisions', in which
agents exchange goods \cite{DL14,DJT,BCMW}, wealth \cite{DT,DMT,BHT13,DLR14}, opinion \cite{T,BMS10,DMPW,MT14,APZ14,DW15} or knowledge \cite{PT14,BLW}. For a general overview on interacting
multi-agent systems and kinetic equations we refer to the book of
Pareschi and Toscani \cite{PT}.

This paper is organised as follows. We introduce a generalization of the kinetic Elo model with variable intrinsic strength 
due to learning in Section~\ref{s:learning}. In Section~\ref{s:fokkerplanck} we derive the corresponding Fokker-Planck 
type equation as the quasi-invariant limit of the Boltzmann type model. Convergence towards steady states of a suitable shifted Fokker-Planck model is analysed in Section~\ref{s:longtime}. We conclude by presenting various numerical simulations of the Boltzmann and the Fokker-Planck type equation in Section~\ref{s:numerics}.

\section{An Elo model with learning}\label{s:learning}

\noindent In this section we introduce an Elo model, in which the rating and the intrinsic strength of the players change in time. 
The dynamics are driven by similar microscopic binary interactions as
in the original model by Jabin and Junca \cite{JJ2015} and Krupp \cite{K2016}. 
We state the specific microscopic interaction rules in each encounter and derive the corresponding limiting Fokker-Planck equation.

\subsection{Kinetic model}

We follow the notation introduced in Section \ref{s:introduction} and denote the individual strength by $\rho$ and the rating by $R$.
If two players with ratings $\Ri$ and $\Rj$ meet, their ratings and strength after the game are given by:
\begin{subequations}	
\begin{align}
\Rit &= \Ri + \var (S_{ij}- b(R_i-R_j)),\\
\Rjt &= \Rj + \var (-S_{ij} - b(R_j-R_i)),\\
\rit &= \ri + \var  h (\rj-\ri) + \eta ,\\
\rjt &= \rj + \var  h (\ri-\rj) + \tilde\eta .
\end{align}\label{e:interactionrules}
\end{subequations}
The interaction rules are motivated by the following considerations:
player ratings change with the outcome of each game (as in the
original model \eqref{e:JJ} proposed by Jabin and Junca \cite{JJ2015}).
The random variable $S_{ij}$ corresponds to the score of the match and depends on the difference in strength of the two players.
We assume that $ S_{ij}$ takes the values $\pm 1$ with an expectation $\langle S_{ij}\rangle= b(\ri-\rj)$.
Note that one could also assume that $S_{ij}$ is continuous, for
example $S_{ij}\in [-1,+1]$. The constant parameter $\var>0$ controls
the speed of adjustment.

The variables $\eta$ and $\tilde{\eta}$ are independent identically
distributed random variables with mean zero and variance $\s$ which model small fluctuations due to day-linked performance
in the mental strength or personal fitness.

The function $h $ describes the learning mechanism. 
%We assume that each player learns in a game, however players with a 
%lower rating benefit more. 
We assume that $h$ takes the following form,
\begin{align}\label{e:h}
%h(\rj-\ri)= \big[\a h_1(\rj-\ri)+\b S_{ij}l(\ri-\rj)\big].
h(\rj-\ri)= \big[\a h_1(\rj-\ri)+\b h_2(\rj-\ri)\big].
\end{align}
The function $h_1$ corresponds to the increase in knowledge or skills because of interactions. We assume that each player learns in a game, however players with a 
lower strength benefit more.  A possible choice for $h_1$, which we shall use throughout this paper, is 
\begin{align}
h_1(\rj-\ri) &= 1+b(\rj-\ri), \label{h1}
\end{align}
where $b$ is given by \eqref{function b}. Note that $b$ is an odd
function. Since $h_1$ is positive, both players are able to learn and improve in each game,
to an extent which depends on the difference in strengths, with a
player with lower strength benefiting more. 

The second function, $h_2$, models a change of strength due to gain or
loss of self-confidence due to winning or being defeated in a game. We
assume that the loss of the stronger player is the same as the gain
for the weaker one. Hence, we choose 
$h_2(\rj-\ri)= S_{ij}l(\rj-\ri) $ 
to be an odd, regular, bounded function which is vanishing at
infinity, where the function $l$ corresponds to the net change of
self-confidence. A possible choice which we adopt in the following corresponds to
\begin{align}
 h_2(\rj-\ri) &= S_{ij}[1-\tanh^2(\rj-\ri)].\label{h2odd}
\end{align}

\begin{figure}
 \includegraphics[width=0.5\textwidth]{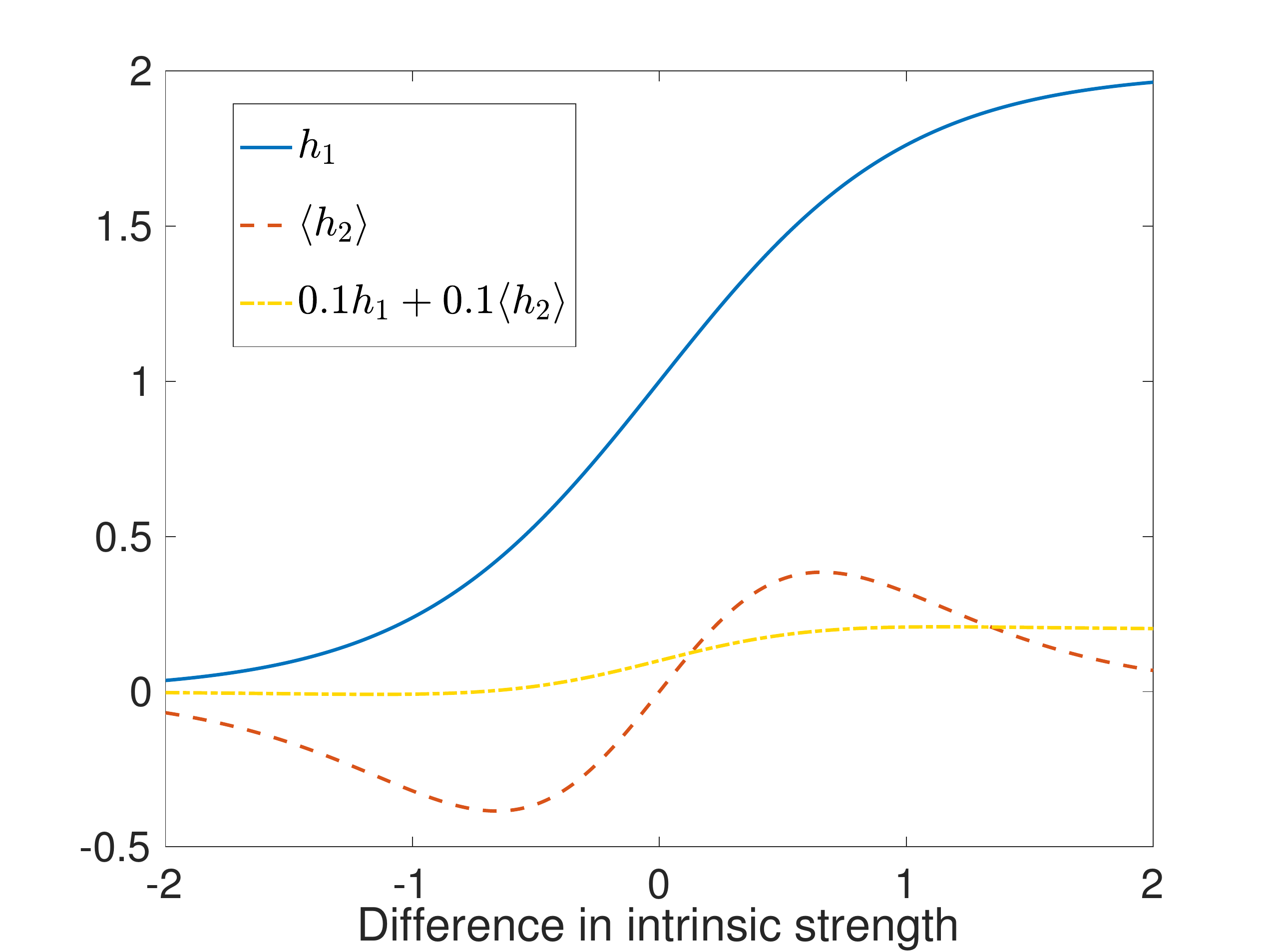}
 \caption{Possible choices of $h_1$ and $h_2$.} \label{f:h}
\end{figure}

Note that the expectation for the learning function function is given by
\be
\langle h(\rj-\ri)\rangle= \big[\a h_1(\rj-\ri)+\b \langle h_2(\rj-\ri)\rangle\big]=\big[\a h_1(\rj-\ri)+\b b(\ri-\rj) (1-\tanh^2(\rj-\ri))\big].\label{learn}
\ee
Figure \ref{f:h} shows the function $h_1$, $\langle h_2 \rangle $ and
$\langle h \rangle$ for the particular choice of $\a = \b = 0.1$ and $c=1$. 
If $\alpha > \beta$ players always improve in strength. In this case the
strength and subsequently the rating will always increase in time. 
We see that, as in the original Elo model, the choices of interaction rules and the function $b(\cdot)$ preserve 
the total value of the rating pointwise and in mean, that is 
$$
\langle\Rit+\Rjt\rangle=\Ri+\Rj.
$$
The evolution of the total strength depends on the choices of the function $h_1$ and $h_2$. 
Note that the function $h_2$ does not affect the total strength since
$$
\langle \rjt+\rjt \rangle-( \rj+\rj )=2\var\a .$$ We see that that the proposed interaction rules result in a net increase of the total knowledge in every interactions. Therefore, we expect to see on overall increase in strength for all times.\\
\revised{The proposed interaction rules are a first step towards  a more realistic modeling. Alternative learning mechanisms, such as the one proposed in the context of knowledge exchange in a large society, see  \cite{BLW}, could be considered in the future. Here the individual with the lower knowledge level assumes the higher level after the interaction, while the stronger one did not gain anything in the encounter. Hence the overall knowledge level is bounded by the maximum initial knowledge level for all times and the distribution of individuals converges to a Delta Dirac at that point. We expect a similar dynamics, if we were to apply that rule instead of \eqref{e:interactionrules}. Developing learning mechanisms, which combine limitations of individual learning with the continuous evolution of the collective knowledge, will be an important aspect of future research developments.}

Now we are able to state the evolution equation for the distribution of players $\fe = \fe(\rho, R, t) $ with respect to their rating $R$ and intrinsic 
strength $\rho$.
\revised{For a fixed number of players,  $N$, the interactions \eqref{e:interactionrules}
induce a discrete-time Markov process  with
$N$-particle joint probability distribution
$P_N(\rho_1,R_1,\rho_2,R_2,\dots,\rho_N, R_N,\tau)$.
One can write a kinetic equation for the one-marginal distribution
function,
$$
P_1(\rho,R,\tau)=\int P_N(\rho, R,\rho_2,R_2,\dots, \rho_N ,R_N,\tau)\,
d\rho_2  dR_2 \cdots  d\rho_NdR_N,
$$
using only one- and two-particle distribution functions \cite{Cerc88,CIP94},
\begin{multline*}
P_1(\rho, R,\tau+1)-P_1(\rho, R,\tau)=\\
\Bigg\langle \frac 1N \Biggl[\int
P_2(\rho_i,R_i, \rho_j, R_j,\tau) w(\Ri-\Rj)\bigl( \delta_0(\rho-\rit,R-\Rit)+\delta_0(\rho-\rjt,R-\Rjt) \bigr)\, d\rho_idR_i d\rho_jdR_j - 2P_1(\rho, R,\tau) \Biggr ]\Biggr\rangle.
\end{multline*}
Here, $\langle\cdot\rangle$ denotes the mean operation with respect to
the random variables $\eta,\tilde\eta$ and the function $w(\cdot)$ corresponds to the interaction rate function which depends on the difference of the ratings.
This process can be continued to give a hierarchy of equations of so-called
BBGKY-type \cite{Cerc88,CIP94}, describing the dynamics of the system of a
large number of interacting agents.
 A standard approximation is to
neglect correlations and assume the factorisation
$$
P_2(\rho_i,R_i, \rho_j, R_j,\tau)=P_1(\rho_i, R_i,\tau)P_1(\rho_j, R_j,\tau).
$$
By scaling time as $t=2\tau/N$ and performing the thermodynamical
limit $N\to\infty$, we can use standard methods of kinetic theory
\cite{Cerc88,CIP94} to show that the time-evolution of the
one-agent distribution function $\fe$ is governed by  the
following  Boltzmann-type equation:}
%The evolution of $\fe$ can be described by the
%following Boltzmann type equation which can be obtained by standard
%methods of kinetic theory:
\be \label{bolt-model-2-weight}
\begin{split}
\frac{d}{dt}\int_{\D} \phi(\ri,\Rj)\fe(\ri,\Ri,t) d\ri d\Ri = \frac{1}{2} \Bigg\langle \int_{\D}\int_{\D} \Big(\phi(\rit,\Rjt)+\phi(\rjt,\Rjt)-\phi(\ri,\Ri)-\phi(\rj,\Rj) \Big)\\
\times\, w(\Ri-\Rj)\fe(\ri,\Ri,t)\fe(\rj,\Rj,t)
\, d\rj d\Rj d\ri d\Ri \Bigg\rangle,\\
\end{split}
\ee
where $\phi(\cdot)$ is a (smooth) test function, with support
$\mathrm{supp}({\phi})\subseteq \D$. The function $w(\cdot)$ corresponds to the interaction rate function which depends on the difference of the ratings. If $w \equiv 1$ we consider a so-called \textit{all-play-all} game.
If $w$ has compact support \revised{ only players with close ratings compete}. Possible choices for $w$ are
\begin{align}
w(\Ri-\Rj) &=e^{\frac{\log 2}{1+(\revisedfurther{\Ri-\Rj})^2}}-1 \text{ or } w(\Ri - \Rj) = \chi_{\{\lvert \Ri - \Rj \rvert \leq c\}}. \label{w}
\end{align}
where $\chi$ denotes the indicator function (or smoothed variants thereof).

 In the following we shall analyse \eqref{bolt-model-2-weight} as well as different asymptotic limits of it.
The presented analysis is based on the following assumptions:
\begin{enumerate}[label=($\mathcal{A}$\arabic*)]
\item \label{a:omega} Let $\Omega = \mathbb{R}^2$ or a bounded Lipschitz domain $\Omega \subset \mathbb{R}^2$. 
 \item \label{a:finit} Let $f_0 \in H^1(\Omega)$ with $f_0 \geq 0$ and compact support. Furthermore we assume that it has mean value zero, and  bounded moments up to order two. Hence
 \begin{align*}
  \int_{\Omega} f_0(\rho,R) \,d\rho dR =1, ~~ \int_{\Omega} R f_0(\rho,R) \,d\rho dR = 0, \text{ and } \int_{\Omega} \rho f_0(\rho,R) \,d\rho dR = 0. \label{ass:finit}
 \end{align*}
 \item The random variables $\eta, \tilde{\eta}$ in
   \eqref{e:interactionrules} have the same distribution, 
   zero mean, $\langle \eta \rangle=0$,  and variance $\sigma_{\eta}^2$.
\item Let the interaction rate function \revisedfurther{$w\ge 0$ be an even} function with 
$w \in C^2(\Omega) \cap L^{\infty}(\Omega)$.
\end{enumerate}

The kinetic Elo model can be formulated on the whole space as well as
on a bounded domain. In reality, the Elo ratings of top chess players
vary between $2000$ to $3000$, which provides evidence for
the assumption of a bounded domain $\Omega$. However, sometimes it is
easier to study the dynamics of models on the whole space, i.e.\
without boundary effects. We will generally work on the bounded domain,
and clearly state where we deviate from this assumption, e.g.\ when we
study the asymptotic behaviour of moments. The second assumption states the necessary regularity assumptions on the initial data, which we shall use in the 
analysis of the moments and the existence proof. 

\subsection{Analysis of the moments}
\label{s:momentsB}
We start by studying basic properties of the Boltzmann type equation
\eqref{bolt-model-2-weight} such as mass conservation and the evolution of the first and second moments with respect to the strength and the ratings.
Throughout this section we consider the problem in the whole space.

\subsubsection*{Conservation of mass:} Setting  $\phi(\ri,\Ri)=1$ in the equation \eqref{bolt-model-2-weight} we see that
$$
\frac{d}{dt} \int_{\R^2} \fe(\ri,\R, t)\,dR d\r=0.
$$
Therefore, the total mass is conserved, that is
\be
\int_{\R^2} \fe(R,\r,t)\,dR d\r=1, \ \text{ for all times } t \geq 0. \label{mass normalisation}
\ee
\subsubsection*{Moments with respect to the rating.}
The $s$-th moment, for  $s\in \mathbb{N}$,  with respect to $\Ri$ is defined as 
$$
\mRi= \int_{\R^2} \Ri \fe(\ri,\Ri,t) \,d\Ri d\ri \text{ and } \ \mathscr{M}_{s,\Ri}(t)=\int_{\R^2} \Ri^s \fe(\ri,\Ri,t) \,d\Ri d\ri,
$$
where $\mRi=\mathscr{M}_{1,\Ri}$.
We choose $\phi(\ri,\Ri)= \Ri$. Due to \ref{a:finit} and the symmetry of $b(\cdot)$ we obtain 
\begin{align*}
\frac{d}{dt} \mRi=&\frac{1}{2}\var\int_{\R^4}  \fe(\ri,\Ri,t)  \fe(\rj,\Rj,t)\times\\
& \times(b(\ri-\rj)-b(\Ri-\Rj)+b(\rj-\ri)-b(\Rj-\Ri)) w(\Ri-\Rj)\,d\Rj d\rj d\Ri d\ri=0.
\end{align*}
Hence the mean value w.r.t.\ the rating is preserved in time and therefore
$$
\mRi=0, \ \text{ for all times } t\geq0.
$$
The evolution of the second moment can be obtained by setting $\phi(\ri,\Ri)= \Ri^2$. We see that
\begin{align*}
\frac{d}{dt} \mathscr{M}_{2,\Ri}(t)=& \frac{1}{2} \int_{\R^4}  \fe(\ri,\Ri,t)   \fe(\rj,\Rj,t) w(\Ri-\Rj)\times\\
&\times\bigg[\var^2\bigg(\big(b(\ri-\rj)-b(\Ri-\Rj)\big)^2+\big(b(\rj-\ri)-b(\Rj-\Ri)\big)^2\bigg)\\
&+2\var \bigg(\Ri (b(\ri-\rj)-b(\Ri-\Rj))+\Rj(b(\rj-\ri)-b(\Rj-\Ri))\bigg)\bigg]d\Rj d\rj d\Ri d\ri.
\end{align*}
The second term in the integral is non-positive and  we obtain the bound
$$
\frac{d}{dt} \mathscr{M}_{2,\Ri}(t)\leq 4 \var^2 \| b\|_{\infty}^2.
$$
Hence, the second moment grows at most linearly and remains bounded for finite times. Note that the integral is negative for $\var$ small enough, which implies a decreasing second moment.

\subsubsection*{Moments  with respect to the strength}
The moments with respect to strength are defined in an analogous way, that is
$$
\mri= \int_{\R^2} \ri f(\ri,\Ri,t) \,d\Ri d\ri  \text{ and } \ \mathscr{M}_{s,\ri}(t)=\int_{\R^2} \ri^s f(\ri,\Ri,t) \,d\Ri d\ri,
$$
for $s \in \mathbb{N}$ and using again $\mri=\mathscr{M}_{1,\ri}$. Since \ref{a:finit} holds, we see that for
$\phi(\ri,\Ri)=\ri$, we have
$$
\frac{d}{dt}\mri=\frac{1}{2}\var \int_{\R^4}\fe(\ri,\Ri,t)\fe(\rj,\Rj,t)w(\Ri-\Rj)[\langle h(\rj-\ri)+h(\ri-\rj) \rangle]\,d\rj d\Rj d\ri d\Ri.
$$
Therefore, 
\begin{align}
-\var\|\langle h \rangle\|_{\infty}    \leq \frac{d}{dt}\mri & \leq\frac{1}{2}\var \int_{\R^4}2\|\langle h \rangle\|_{\infty} \fe(\ri,\Ri,t)\fe(\rj,\Rj,t) d\rj d\Rj d\ri d\Ri
\leq\var \|\langle h \rangle\|_{\infty},\label{1-mom-bolt}
\end{align}
which implies that the mean value is bounded for all times $t \in
[0,T]$ and that $|\mri|$ grows at most linearly in time if $h(\cdot)$
is bounded.
If we consider the specific interaction rules \eqref{h1}-\eqref{w}, we obtain
$$
\frac{d}{dt}\mri = \var \a \int_{\R^4}w(\Ri-\Rj)\fe(\ri,\Ri,t)\fe(\rj,\Rj,t) d\rj d\Rj d\ri d\Ri\le\var\a,
$$
with equality holding in the ``all-play-all'' case $w=1$.
The evolution of the second moment $\mathscr{M}_{2,\ri}$ can be computed by setting $\phi(\ri,\Ri)=\ri^2$. We see that
\begin{align}
\begin{split}
\frac{d}{dt}\mathscr{M}_{2,\ri}(t)=&\frac{1}{2} \int_{\R^4}\big (\var^2[\langle h(\rj-\ri)^2\rangle+\langle h(\ri-\rj)^2\rangle]+ 2\var[\ri\langle h(\rj-\ri)\rangle+\rj\langle h(\ri-\rj)\rangle]\\
& \hspace*{3cm} + 2 \s(\var)\big )w(\Ri-\Rj)\fe(\ri,\Ri,t)\fe(\rj,\Rj,t)\,d\rj d\Rj d\ri d\Ri \\
\leq & \var^2\|\langle h^2 \rangle\|_{\infty}+\s(\var)+4\var|\mri|.
\end{split}
\label{2-mom-r}
\end{align}
If $h(\cdot)$ is bounded the second moment grows at most at polynomial rate. Since the second moment of $f_0$ is bounded (see assumption
\ref{a:finit}), it remains finite for all times $t\in [0,T]$.

\section{The Fokker-Planck limit}\label{s:fokkerplanck}

\noindent In the last section we analysed the evolution of moments to the Boltzmann type equation \eqref{bolt-model-2-weight}. However, it is often more useful to study  the dynamics
of simplified models (generally of Fokker-Planck type), which can be derived in particular asymptotic limits. These asymptotics 
provide a good approximation of the stationary profiles of the kinetic equation. In what follows we consider the so-called \textit{quasi-invariant} limit, in which
diffusion and the outcome of the game influence the long-time
dynamics. More specifically, we consider the limit
\begin{align*}
 \gamma \rightarrow 0,\, \sigma_{\eta} \rightarrow 0 \text{ such that
  } \frac{\sigma_{\eta} ^2}{\gamma} =: \sigma^2 \text{ is
  kept fixed}.
\end{align*}

In Appendix \ref{s:appendix} we derive the following Fokker-Planck limit:
The differential form of \eqref{integral-FP-model 2} is given by
(writing $t$ instead of $\tau$)
\be \label{differential-FP-model}
\frac{\partial f(\r,R, t)}{\partial t}= - \deR \left(a[f]f(\r,R,
  t)\right)-\dr \left(c[f]f(\r,R, t)\right) +\frac{\s}{2} d[f]  \drr
f(\r,R, t) \text{ in} \ \D\times (0,T),
\ee
where
\begin{align*} 
&a[f]=a[f](\r,R, t)=\int_{\R^2}w(R-\Rj) (b(\r-\rj)-b(R-\Rj))f(\rj,\Rj,t)\,d\rj d\Rj ,\\
&c[f]=c[f](\r,R, t)=\int_{\R^2} w(R-\Rj)\big( \a h_1(\rj-\r)+\b
  \langle h_2(\rj-\r)\rangle\big)f(\rj,\Rj,t)\,d\rj d\Rj, \\
&d[f]=d[f] (R,t)= \int_{\R^2}w(R-\Rj) f(\rj,\Rj,t)\,d\rj d\Rj.
\end{align*}
We consider equation \eqref{differential-FP-model} with initial datum $f_0$ satisfying assumption \ref{a:finit} in the following. 
Note that \eqref{differential-FP-model} includes the nonlocal operator $a[f]$,
corresponding to the change of the ratings, similar as in the
Fokker-Planck equations \eqref{e:originalboltzmann} and
\eqref{e:boltzmannkrupp} obtained in \cite{JJ2015} and \cite{K2016}, respectively.
The nonlocal operator $c[f]$  in the transport terms corresponds to
the change of the individual strengths while the 
operator $d[f]$ describes the fluctuations of the individual strength due to encounters.

\subsection{Qualitative properties of the Fokker-Planck equation}

We continue by discussing qualitative properties of the Fokker-Planck equation \eqref{differential-FP-model}. We shall see that several properties, which
we observed for the Boltzmann type equation \eqref{bolt-model-2-weight}, can be transferred.\\

\noindent \textit{Conservation of mass and positivity of solution:} Due to mass conservation and \ref{a:finit} we have that
$$
\int_{\R^2} f(\r,R,t)\, d\r dR=\int_{\R^2} f_0(\r,R) \,d\r dR=1 \text{ for
  all } t\ge 0.
$$
Using similar arguments as in \cite{TT}, we can directly prove that
the Fokker-Planck equation maintains the positivity of the
solution. Let $v_m(\title{t})= (\r_m(t), R_m(t))$ denote the minimum, which is obtained at time
$\tilde{t}$. Clearly, if at certain time  $\tilde{t} \geq 0$ the
function equals zero, i.e.\ $f(\r,R,\tilde{t})=0$, this point
is a stationary point or a local minimum, hence
$$
\deR f(v_m,\tilde{t})=0, \quad \dr f(v_m,\tilde{t})=0,\quad
\deRR f(v_m,\tilde{t})\geq 0, \quad \drr f(v_m,\tilde{t})\geq 0.
$$
Evaluating \eqref{differential-FP-model} in $(v_m,\tilde{t})$ gives
\begin{align*}
\dt f(v_m,\tilde{t})=& f(v_m,\tilde{t})\big(-\deR a[f](v_m,\tilde{t})-\dr c[f](v_m,\tilde{t})\big)\\
&-a[f](v_m,\tilde{t})\deR f(v_m,\tilde{t})- c[f](v_m,\tilde{t}) \der f(v_m,\tilde{t})
+\frac{\s}{2}(v_m,\tilde{t}) d[f]\drr \left( f(v_m,\tilde{t})\right)\ge 0,
\end{align*}
which implies that the function $f$ is non-decreasing in time and cannot assume negative
values.\\

\noindent \textit{Evolution of the moments:} We now consider the evolution of the moments of the solution of
\eqref{differential-FP-model} using the interaction rules \eqref{h1} and \eqref{h2odd}.
Similar calculations as in Section~\ref{s:momentsB} confirm the
expected behaviour ---due to the continuous increase in strength in
each game the system does not converge to a steady state and therefore
the respective mean of the solution is non-decreasing in
time. Summarising the results, we have
\begin{align}
\dt \int_{\R^2} R f(\r,R,t)\,dR d \r &=0 \label{mean R FP}\\
\begin{split}
\dt\int_{\R^2} \r f(\r,R,t)\,dR d \r &= \a \int_{\R^2} c[f] f(\r,R,t)\,dR d \r\\
								&=\a \int_{\R^4}w(R-\Rj)f(\r,R,t)f(\rj,\Rj,t)\,d \rj d \Rj d\r dR.\label{mean rho FP}
								\end{split}
\end{align}

The previous results confirm that due to the continuous increase in
strength in each game, rating and skills tend to become increasingly
distant from each other. Therefore, we adopt an idea by Krupp \cite{K2016} and study the evolution of a suitably shifted problem instead. We define
\be
g(\r,R,t)=f(\r+H(\r,R,t),R,t) \label{scaling r},
\ee
where the scaling function $H$ is given by 
\be \label{Scaling function}
\frac{\partial H(\r,R,t)}{\partial t}=\int_{\R^2}\a w(R-\Rj)f(\rj,\Rj,t)\,d \rj d \Rj=\a d[f] .
\ee
This scaling ensures that the mean value is preserved in time. The corresponding evolution equation for $g(\r,R,t)$ is given by
%\be \label{FP-scaling r}
%\frac{\partial g(\r,R, t)}{\partial t}= - \deR (a[g]g(\r,R, t))-\dr (c[g]g(\r,R, t)) +\frac{\s}{2}d[g]\drr (g(\r,R, t))+ \int_{\R^2}\a w(R-\Rj)f(\rj,\Rj,t)d \rj d \Rj \dr g(\r,R, t).
%\ee
%or equivalently by
\begin{equation*} %\label{FP-scaling r-tilde}
\frac{\partial g(\r,R, t)}{\partial t}= - \deR (a[g]g(\r,R, t))-\dr (\ct [g]g(\r,R, t)) +\frac{\s}{2}  d[g] \drr g(\r,R, t),
\end{equation*}
where
$$
\ct [g]=\ct [g](\r,R, t)=\int_{\R^2} \big( \a b(\rj-\r)+\b \langle h_2(\rj-\r)\rangle\big)w(R-\Rj) g(\rj,\Rj,t)\,d\rj d\Rj.
$$
Now, the mean value of $g(\r,R, t)$ is constant w.r.t. both $R$ and $\r$ and we can normalize 
$$
\int_{\R^2}R g(\r,R,t) \,d \r dR=0, \text{ and } \int_{\R^2}\r g(\r,R,t) \,d \r dR=0.
$$
In a general setting it is not possible to compute scaling function
explicitly. However, in `all-meet-all' tournaments, that is $w(R-\Rj)=1$, and in case of the specific interaction
rules \eqref{h1}-\eqref{h2odd},
we obtain that $$H(\r,R,t)=\a t.$$
Therefore, in the rest of this paper, we consider the following problem on a bounded domain $\D\subset\R^2$, with no-flux boundary condition
\begin{subequations}\label{e:FPg}
\begin{align} 
\frac{\partial g(\r,R, t)}{\partial t}&= - \deR (a[g]g(\r,R, t))-\dr
                                        (\ct [g]g(\r,R, t))
                                        +\frac{\s}{2} d[g] \drr 
                                        g(\r,R, t) && \text{in} \
                                                        \D\times (0,T),\label{FP-scaling r-tilde}
\\
\frac{\partial}{\partial{\nu}} g&=0 & &\text{on} \ \partial \D, \label{noflux BC g}\\
 g(\r,R,0)& =g_0(\r,R) & &\text{in} \ \D. \label{initial data g}
\end{align}
\end{subequations}
\revised{Here $\nu$ denotes the unit outer normal vector.}
Note that the existence of solutions to \eqref{FP-scaling r-tilde} on the whole domain is more involved, since we would need to
prove that the solution decays sufficiently as $R$ and $\r$ tend to infinity. Therefore,
we consider the equation on a bounded domain only.

\subsection{Analysis of the Fokker-Planck equation}\label{s:analysis}
In the section we prove existence of weak solutions to \eqref{e:FPg}. The main result reads as follows.
\begin{theorem}
Let \ref{a:omega} be satisfied, $g_0 \in H^1(\D)$ and $0\le g_0 \le M_0$ for some
  $M_0>0$ and assume $h_1$, $\langle h_2\rangle$, $b$ $\in L^\infty(\D)\cap
  C^2(\D)$. Then there exists a weak solution
$g\in\LDt\HD\cap \HDt\HDd$ to \eqref{FP-scaling r-tilde}--\eqref{initial data g}, satisfying $0\le g
\le M_0 e^{\lambda t}$ for all $(\r,R) \in \D$, $t>0$, with a constant
$\lambda>0$ depending on the functions $h_1, \langle h_2\rangle, b$ and $w$.
\end{theorem}

The presented existence proof was adapted from a similar argument for a nonlinear Fokker-Planck equation
describing the dynamics of agents in an economic market, see \cite{DJT}. However, equation \eqref{FP-scaling r-tilde}
has an additional nonlinearity in the derivative w.r.t. the rating $R$. We divide the proof in several
steps for the ease of presentation. In \textit{Step 0} we regularize the non-linear Fokker Planck equation \eqref{FP-scaling r-tilde}
by adding a Laplace operator with small diffusivity $\mu \geq 0$. We linearise the equation in \textit{Step
  1} and show existence of a unique solution for this problem. In \textit{Step 2} we derive the necessary $L^{\infty}$ estimates 
  to use Leray-Schauder's fixed point theorem and show existence of solutions to the nonlinear regularised problem. In \textit{Step 3}
  we present additional $H^1$ estimates, which allow us to pass to the limit $\mu \rightarrow 0$  in \textit{Step 4}.

\begin{proof}
\noindent\textit{Step 0: the regularised problem.} For $M>0$, let us denote by  $g_M=\max\{0,\min\{g,M\}\}$ and define
\begin{align*}
K_M[g]&=\int_{\D} [\a h_1(\rj-\r)+\b \langle h_2(\rj-\r)\rangle]w(R-\Rj)g_M(\rj,\Rj,t)\,d\rj d\Rj,  \\
L_M[g]&=\int_{\D} [b(\r-\rj)-b(R-\Rj)]w(R-\Rj)g_M(\rj,\Rj) \,d\rj d\Rj.
\end{align*}
Next we consider the regularised non linear problem for $0<\mu<1$,
\begin{subequations}\label{e:FPgreg}
\begin{align}
\begin{split}
\dt \gmu= - \deR (\A \gmu(\r,R, t))&-\dr (\K \gmu(\r,R, t)) \\
&+\frac{\s}{2}d[\gmu] \drr (  \gmu(\r,R, t))+\mu \Delta (\gmu(\r,R, t)) \ \text{in} \
                                                        \D\times (0,T),\label{eq15-w}
                                                        \end{split}
\end{align}
with boundary and initial conditions given by
\begin{align}
\frac{\partial}{ \partial \nu} \gmu=0 \ \text{on} \ \partial \D, \text{ and } \gmu(\r,R,0)=g_0  \
 \text{on} \ \D. \label{eq16-w}
\end{align}
\end{subequations}
The weak formulation of \eqref{e:FPgreg} is given by
\be
\int_0^T \Bigl\langle \dt \gmu,v \Bigr\rangle \,dt= \int_0^T \int_{\D} \bigg(\A \gmu\deR v+\K \gmu \dr v-\frac{\s}{2}d[\gmu]\dr \gmu \dr v-\mu \deR \gmu \deR v\bigg)\,dR d\r dt \label{eq17-w},
\ee
where $\langle \cdot,\cdot\rangle$ is the dual product between $\HD$
and  $\HDd$ and $v \in H^1(\Omega)$.

%%%%

\noindent\textit{Step 1: solution of the linearised regularised problem.} Next we want to apply Leray-Schauder's fixed point theorem. Let $\gt \in L^2(0,T;L^2(\D))$, $\theta\in [0,1]$
and $g^+ = \max(g,0)$. 
  We introduce the operators $A:\HD\times\HD\rightarrow \R$ and $F:\HD\rightarrow \R$:
\begin{align}
A(\gmu,v)&=\int_{\D} \mu\bigg( \deR \gmu \deR v  + \dr \gmu \dr v\bigg)\,dR d\r , \label{eq-18-w}\\
F(v)&=\theta \int_{\D} \bigg( L_M[\gt] \gt^+ \deR v  +K_M[\gt] \gt^+ \dr v-\sm d[\gt] \dr  \gt^+ \dr v\bigg) \,dR d\r. \label{eq-19-w} 
\end{align}
The operator $A(\cdot,\cdot)$ is bilinear and continuous on $\HD \times \HD$.
%\begin{align*}
%|a(\gmu,v)|&\leq \sm\int_\D \left|\dr \gmu\right|\left|\dr v\right|+\mu \left|\deR \gmu\right|\left|\deR v\right|\\
%&\leq \sm\int_\D \frac{1}{2}\left(\left|\dr \gmu\right|^2+\left|\dr v\right|^2\right)+\mu \int_\D \frac{1}{2}\left(\left|\deR \gmu\right|^2+\left|\deR v\right|^2\right)\\ 
%&\leq \frac{1}{2}\int_\D \left(\sm+\mu\right)\left(\|\gmu\|^2_{\HD}+\|v\|^2_{\HD}\right).
%\end{align*}
The quantities $|K_M[\gt]|$ and $|L_M[\gt]|$ are bounded (because of
the assumption made on $h_1, \langle h_2 \rangle $ and $b$), therefore $F$ is continuous in $\HD$.
%\begin{align*}
%|F(v)|&\leq \n \int_\D \left(\left|K_M[\gt]\right|\left|\gt^+\dr v\right|+\left|L_M[\gt]\%right|\left|\gt^+\deR v\right|\right)\\
%&\leq C\left(\norm{\gt^+}{\LD}^2+\norm{v}{\HD}^2\right).
%\end{align*}
Because of Poincar\'{e}'s inequality, for some constant $C_1$ and $C_2$
\begin{align*}
A(\gmu,\gmu)&= \mu \int_\D \Big(\Big|\dr \gmu\Big|^2 +\Big|\deR \gmu\Big|^2\Big) \,dR d\r\geq C_1\mu \norm{\gmu}{H^1(\D)}-C_2 \norm{\gmu}{2}.
\end{align*}
By corollary $23.26$ in \cite{Z}, there exists a unique solution  $\gmu\in \LDt{\HD} \cap \HDt{\HDd}$ \nolinebreak to 
\be
\Bigl\langle \dt \gmu,v \Bigr\rangle+ A(\gmu,v)=F(v), \ t>0, \ \gmu(0)=\theta g_0. \label{eq-20-w}
\ee
This defines the fixed-point operator $V : \LDt{\LD}\times
[0,1]\rightarrow \LDt\LD $, $(\gt,\theta)\mapsto  V(\gt,\theta)=\gmu$, where $\gmu$ solves $\eqref{eq-20-w}$. This operator satisfies $V(\gt,0)=0$. %Indeed for $\theta=0$, equation $\eqref{eq-20}$ becomes $\langle \dt \gmu,v\rangle+ a(\gmu,v)=0 $, $\gmu(0)=0$. \\
Standard arguments, including Galerkin's method and estimates on $\norm{\dt \gmu}{\LDt{\HDd}}$, show that the operator $V$ is continuous (with
constants depending on the regularisation parameter $\mu$).
%For $\theta=0$, $\norm{\gmu}{\LDt\LD}=0$. For $\theta\neq0$, considering that \linebreak $\norm{\gmu}{\LDt\LD}\leq \norm{\gmu}{\HDt\LD}$ we can use the \eqref{eq-20} with $v=\gmu$ to compute the evolution of the norm of $\gmu$:
%$$
%2 \frac{d}{dt}\norm{\gmu}{\HDt\LD}^2=-A(\gmu,\gmu)+F(\gmu)\leq C_1\norm{\gmu}{H^1(\D)}^2+C_2\norm{\gt}{\LD}^2,
%$$
%then, integrating in time, it results
%$$
%\norm{\gmu}{\HDt\LD}\leq e^{C_1 T} C_2 \int_0^T \norm{\gt}{\LD}^2e^{C_1 t} dt= C(T) \norm{\gt}{\LDt\LD}^2.
%$$
%We can estimate $\norm{\dt \gmu}{\LDt{\HDd}}$, using the norm of operators $\norm{\dt \gmu}{\HDd}= \sup_{\norm{v}{\HD}=1} |\langle \dt \gmu, v\rangle|$.\\
% For a suitable $C \geq(\norm{\deR \A}{\infty})^\frac{1}{2}+(\norm{\dr \K}{\infty})^\frac{1}{2}+\sm+1$
%\begin{align*}
%|\langle \dt \gmu,v\rangle|&\leq \norm{\A}{\infty}\int_\D \big(\gmu^2+|\deR v|^2\big)+\norm{\K}{\infty}\int_\D \big(\gmu^2+|\dr v|^2\big)\\
%&+\sm \norm{d[\gmu]}{\infty}\int_\D \big(|\dr \gmu|^2+|\dr v|^2\big)+\mu\int_\D\big( |\nabla \gmu|^2+|\nabla v|^2\big)\\
%&\leq C(\norm{\gmu}{H^1(\D)})\norm{v}{H^1(\D)}
%\end{align*}
%It implies that for a suitable $C$ that doesn't depend on $\mu$
%\be 
%\norm{\dt \gmu}{\LDt{\HDd}}\leq C. 
%\ee
The operator is also compact, because $\LDt\HD\cap\HDt{\HDd}$ is compactly embedded in $\LDt{\LD}$, see \cite{Sim}.
In order to apply the fixed-point theorem of Leray-Schauder, we need
to show uniform estimates.

%%%%

\noindent\textit{Step 2: uniform $L^\infty$ bound \& existence of a
  fixed point.} We start by proving upper and lower bounds for the function $\gmu$. Let $\gmu$ be a fixed point of $V(\cdot,\theta)$, i.e.\
$\gmu$ solves \eqref{eq-20-w} with $\gt=\gmu$, and $\theta\in[0,1]$.\\
For a lower bound, choosing $v=\gmu^-=\min\{0,\gmu\}\in \LDt\HD$ as test function in \eqref{eq-20-w} and integrating in time, we obtain
$$
\frac{d}{dt}\norm{\gmu^-}{\LD}^2=-2A(\gmu,\gmu^-)\leq -C_1\norm{\gmu^-}{2}^2\leq 0.
$$  
This shows that if $\gmu(0)^-=0$, then $\gmu(t)^-=0$ for all $t>0$. Hence,
in all previous computations and in \eqref{eq-18-w}-\eqref{eq-19-w}, we
can replace $\gmu^+$ with $\gmu$.

Now we show an upper bound. Let $\gm=(\gmu-M)^+$, where
$M=M_0e^{\l t}$, for some $\l>0$ to be determined below. We choose $v=\gm \in \LDt\HD$ as
test function in \eqref{eq17-w}. By assumption, $g_0\leq M_0$,
i.e.\ $\gm(0)=(g_0-M_0)^+=0$. We note that 
$\dt M=\l M$ and $\frac{1}{2}\dr
(\gm^2)=(\gmu-M)\dr \gm$. Then  
\begin{align*}
\frac{1}{2} \int_\D \gm(t)^2 \,dR d\r&=\int_0^t \left[- \l \int_\D M \gm \, \,dR d\r -A(\gmu,\gm)+F(\gm) \, \right]ds\\
&=\int_0^t \sm \int_\D d[\gmu]\dr ((\gmu-M)+ M)\dr \gm \,dR d\r -\mu \int_\D |\nabla \gm|^2\,dR d\r +\theta(I+J)\,ds\\
&\le \int_0^t\theta (I+J) \,ds,
\end{align*} 
where 
$I=\int_\D \A \gmu \deR \gm \,dR d\r $ and $ J=\int_\D \K \gmu \dr \gm \,dR d\r .$
Let us consider $I$ and $J$ separately:
\begin{align*}
I&=\int_\D \A (\gmu-M)\deR \gm \,dRd\r+\int_\D\A M \deR \gm \,dRd\r\\
&=-\frac{1}{2} \int_\D \deR [\A]\gm^2\, dRd\r- \int_\D \deR [\A]M \gm \,dRd\r \\
J&=\int_\D \K (\gmu-M)\dr \gm\, dRd\r +\int_\D\A M \dr \gm \,dR d\r\\
&=-\frac{1}{2} \int_\D \dr [\K]\gm^2\, dRd\r- \int_\D \dr [\K]M \gm \,dR d\r.
\end{align*}
The assumptions on $h_1$, $\langle h_2\rangle$ and $b$ ensure that $\deR [\A]$ and $\dr [\K]$ are bounded. Hence
\begin{align*}
\frac{1}{2}\int_\D \gm^2 \,dR d\r&=\int_\D \Big(\dt \gm\Big)\gm \,dR d\r \\
&\leq C(\A,\K)\int_\D \gm^2 \,dR d\r + (C(\A,\K)-\l)\int_\D M \gm \,dR d\r.
\end{align*}
Choosing $\l$ large enough and using Gronwall's lemma, we obtain
$$
\int_\D \gm(t)^2\, dR d\r \leq \int_\D \gm(0)^2 \exp[2C(\A,\K) t  ]\, dR d\r=0.
$$
Therefore $\gm(t)=0$ for all $t>0$, which implies  $\gmu(t)\leq M$ for all $t>0$. 
This allows us to replace $\A$ with $a[\gmu]$ and $\K$ with $\ct[\gmu]$ in
\eqref{eq17-w}. The uniform $L^\infty$ bound provides the necessary bound for
the fixed-point operator in $\LDt\LD$. This implies existence of a weak solution to \eqref{eq17-w}.

%%%%

\noindent\textit{Step 3: uniform $H^1$ bound.}
Our aim is to derive an $H^1$ bound which is independent of $\mu$.
Choosing $v=\gmu$ in \eqref{eq17-w} with $t$ instead of $T$, we obtain
\begin{align*}
\frac{1}{2}\frac{d}{dt}\int_\D \gmu(t)^2 \,dR d\r &=\int_\D a[\gmu]\gmu\deR \gmu \,dR d\r +\int_\D \ct[\gmu]\gmu\dr g\,dR d\r\\
& \phantom{=}- \int_\D \Big(\sm d[\gmu]+\mu\Big)\Bigl|\dr \gmu\Bigr|^2\,dR d\r-\mu\int_\D \Bigl|\deR \gmu\Bigr|^2\,dR d\r\\
&= -\frac{1}{2}\int_\D \deR a[\gmu] \gmu^2\,dR d\r-\frac{1}{2}\int_\D \dr \ct[\gmu] \gmu^2\,dR d\r- \int_\D \Bigl(\sm d[\gmu]+\mu\Bigr)\Bigl|\dr \gmu\Bigr|^2\,dR d\r\\
&\phantom{=}-\mu\int_\D \Bigl|\deR \gmu\Bigr|^2\,dR d\r.
\end{align*}
Because of the assumptions on $h_1,$ $\langle h_2 \rangle$ and $b$ we have that $\left|-\frac{1}{2}\left(\deR a[\gmu]+\dr \ct[\gmu]\right)\right|<C$. Therefore, we can rewrite the above estimate as
\begin{align}\label{eq21-w}
\begin{split}
\frac{1}{2}\int_\D \gmu(t)^2\,dR d\r&+\int_0^t\bigg[ \int_\D \Bigl(\sm d[\gmu]+\mu\Bigr)\Bigl|\dr \gmu\Bigr|^2\,dR d\r+\mu\int_\D \Bigl|\deR \gmu\Bigr|^2\,dR d\r\bigg]ds\\
&\leq C \int_0^t\int_\D \gmu(t)^2 \,dR d\r dt +\frac{1}{2}\int_\D g(0)^2 \,dR d\r.
\end{split}
\end{align}
Using Gronwall's lemma, the previous estimate guarantees (independent by $\mu$) estimates for $\gmu(t)$, i.e.  
\begin{equation*}
\norm{\gmu}{L^\infty(0,T;\LD)}\leq C.
\end{equation*}
However, this does not ensure an (independent of $\mu$) estimate for
$\deR \gmu$ and $\dr \gmu$. 
 In order to obtain it, we differentiate \eqref{eq15-w} with respect to $R$ and $\rho$ in the sense of distributions.
 This gives us estimates for $y:=\deR \gmu$ and $z:=\dr \gmu$. We obtain
\be
\dt y= -\deR \big( d[\gmu]\gmu+a[\gmu] y\big)-\dr (\ct[\gmu]y)+\sm \drr y + \var
\deRR y \quad \text{in} \ \D\times (0,T).
\label{eq-22}
\ee
Due to no-flux boundary condition \eqref{noflux BC g}, equation \eqref{eq-22} is complemented with 
$$
\frac{\partial}{\partial \nu_R} y(\r,R,t)=0  \quad \text{on} \ \partial \Omega,
$$
where $\nu_R$ is the component w.r.t. variable $R$ of the normal vector $\nu$ to $\D$.
Furthermore $y(\r,R,0)=\deR g_0(\r,R)$. Choosing $v\in \LDt{H^1_0(\D)}$ and setting $d'[\gmu]=\deR d[\gmu]$, $\ct_R[\gmu]=\deR \ct[\gmu]$ and  $a_R[\gmu]=\deR a[\gmu]$, 
we obtain the weak formulation of equation \eqref{eq-22}:
\begin{multline}
\int_0^T\Bigl\langle\dt y,v\Bigr\rangle\, ds=\int_0^T \int_\D \Bigl(a_R[\gmu]\gmu
\deR v +a[\gmu]y \deR v+\ct_R[\gmu]\gmu\dr v+\ct[\gmu]y\dr v \label{eq22weak-w}\\
-\sm \dr \bigl(d'[\gmu]\gmu+d[\gmu]  y\bigr)\dr v- \mu\Bigl( \dr y \dr v+\deR y \deR v\Bigr)\Bigr)\,dR d\r ds.
\end{multline}
We introduce the operators 
\begin{align*}
 B_y(y,v)&=\int_\D -a[\gmu]y \deR v-\ct[\gmu]y\dr v+\sm d[\gmu] \dr y\dr v + \mu \Bigl( \dr y \dr v+\deR y \deR v\Bigr)\,dR d\r \\
 G_y(v)&= \int_\D \ct_R[\gmu]\gmu\dr v +a_R[\gmu]\gmu \deR v-\sm d'[\gmu]\dr \gmu \dr v \,dR d\r.
\end{align*}
Both operators $B_y: \LDt{H^1_0(\D)}\times  \LDt{H^1_0(\D)}\rightarrow \R$ and $G_y:\LDt{H^1_0(\D)}\rightarrow \R$ are
linear and continuous. Garding's inequality implies
\begin{align*}
B_y(y,y)&=\int_\D \mu |\nabla y|^2 \,dR d\r +\frac{1}{2}\int_\D (\ct_R[\gmu]+a_R[\gmu])y^2 d\r dR+\sm\int_\D d[\gmu] \Bigl|\dr y\Bigr|^2 \,dR d\r\\
&\geq \mu \norm{y}{H^1(\D)}^2-\Bigl(\mu+\frac{1}{2}\norm{ a[\gmu]}{\infty}+\frac{1}{2}\norm{ \ct[\gmu]}{\infty}\Bigr)\norm{y}{2}^2.
\end{align*}
Then corollary $23.26$ in \cite{Z} gives existence of a unique solution $y \in \LDt{H^1_0(\D)} \cap \HDt{H^{-1}(\D)}$ to 
\be
\Bigl\langle \dt y,v\Bigr\rangle+ B_y(y,v)=G_y(v), \ t>0, \ y(0)= y_0. 
\ee
Choosing $v=y$ in \eqref{eq22weak-w}, we obtain (using Young's and Gardin's inequality)
\begin{align*}
\frac{1}{2}\frac{d}{dt} &\int_\D y(t)^2 \,dR d\r = -B_y(y,y)+G_y(y)\\
\leq&-\mu \norm{y}{H^1(\D)}^2+C\norm{y}{2}^2+\frac{1}{2}\Big(\norm{\deRR a[\gmu]}{\infty}+\norm{\dr\Bigl(\deR \ct[\gmu]\Bigr)}{\infty}\Big)\int_\D \gmu^2+y^2+\Bigl|\dr \gmu\Bigr|^2\,dRd\r\\
&-\sm\int_\D d'[\gmu] \dr \gmu \dr y \,dR d\r.
\end{align*}
Considering the last integral, we calculate
\begin{align*}
-\sm\int_\D d'[\gmu] \dr \gmu \dr y  \,dR d\r&=-\sm\int_\D d' [\gmu] \dr \gmu\dr \Bigl(\deR \gmu\Bigr) \,dR d\r\\
&= \sm \int_\D \deR d'[\gmu] \Bigl|\dr \gmu\Bigr|^2 \,dR d\r+\sm \int_\D d'[\gmu] \deR\Bigl (\dr \gmu\Bigr)\dr \gmu \,dR d\r,
\end{align*}
and therefore,
$$
-\sm\int_\D d'[\gmu] \dr \gmu \dr y \,dR d\r=\frac{\s}{4}\int_\D \deR d'[\gmu] \Bigl|\dr \gmu\Bigr|^2 \,dR d\r.
$$
This gives us the following estimate for $\norm{y}{\LD}$ (with a constant depending on $a[\gmu]$, $\ct[\gmu]$ and their derivatives)
\be 
\int_\D y(t)^2 \,dR d\r \leq \int_\D h(0)^2\, dR d\r +
C\int_0^t\int_\D y^2+\gmu^2+\Bigl|\dr \gmu\Bigr|^2\,dR d\r ds.\label{eq-24-w}
\ee

We use similar arguments for $z=\dr \gmu$.
For a suitable $C$, which depends on $a[\gmu]$, $\ct[\gmu]$, $d[\gmu]$ and their derivatives (but not on $\mu$), we obtain an estimate for the $L^2$ norm of $z$:
\be 
\int_\D z(t)^2\,dRd\r \leq \int_\D h(0)^2\, dRd\r+
C\int_0^t\int_\D z^2+\gmu^2+\Bigl|\deR \gmu\Bigr|^2 \,dRd\r ds.\label{eq-24-w-z}
\ee
We add  \eqref{eq21-w}, \eqref{eq-24-w} and \eqref{eq-24-w-z} to obtain
\begin{multline}
\int_\D \gmu(\r,R,t)^2+y(\r,R,t)^2+z(\r,R,t)^2\, dRd\r+\sm\int_0^t  \int_\D z(\r,R,s)^2\, dR d\r ds \\
\leq C \int_0^t\int_\D y(\r,R,s)^2+\gmu(\r,R,s)^2+z(\r,R,s)^2\, dR d\r ds+
\int_\D g(\r,R,0)^2+y(\r,R,0)^2+z(\r,R,0)^2\, dR d\r,
\end{multline}
where $C$ does not depend on $\mu$. Using Gronwall's lemma gives the following estimates (independent of $\mu$)
\begin{align}
&\norm{\gmu}{L^\infty(0,T;\LD)}\leq C, \quad \norm{\dr \gmu}{L^\infty(0,T;\LD)}\leq C, \quad \norm{\deR \gmu}{L^\infty(0,T;\LD)}\leq C. \label{eq25-deRg-w}
\end{align}

%%%%

\noindent \textit{Step 4: The limit $\mu \rightarrow 0$.}
Let $\gmu$ solution of \eqref{eq15-w}-\eqref{eq16-w} with $L[\gmu]=a[\gmu]$ and $K[\gmu]=\ct[\gmu]$. We can estimate $\norm{\dt \gmu}{\LDt\HDd}$, using the norm of operators $\norm{\dt \gmu}{\HDd}= \sup_{\norm{v}{\HD}=1} |\langle \dt \gmu, v\rangle|$.\\
 For a suitable $C \geq(\norm{\deR a[g]}{\infty})^\frac{1}{2}+(\norm{\dr \ct[g]}{\infty})^\frac{1}{2}+\sm+1$, we obtain
\begin{align*}
\Bigl|\Bigl\langle \dt \gmu,v \Bigl\rangle\Bigr| &\leq \norm{a[\gmu]}{\infty}\int_\D \Big(\gmu^2+\Bigl|\deR v\Bigr|^2\Big)\,dRd\r+\norm{\ct[\gmu]}{\infty}\int_\D \Big(\gmu^2+\Bigl|\dr v\Bigr|^2\Big)\,dRd\r\\
&\phantom{\leq}+\sm \norm{d[\gmu]}{\infty}\int_\D \Bigl|\dr \gmu\Bigr|^2+\Bigl|\dr v\Bigr|^2\,dRd\r+\mu\int_\D |\nabla \gmu|^2+|\nabla v|^2\,dRd\r\\
&\leq C(\norm{\gmu}{H^1(\D)})\norm{v}{H^1(\D)}.
\end{align*}
This implies 
\be \label{eq26-w}
\norm{\dt \gmu}{\LDt\HDd}\leq C \text{ and } \int_0^T \norm{\gmu}{\HD}^2 dt=C\norm{\gmu}{\LDt\HD}\leq C, 
\ee
where $C$ does not depend on $\mu$.
% and the dimension of the domain $\D$.
Estimates \eqref{eq25-deRg-w} and \eqref{eq26-w} allow us
to apply Aubin-Lions lemma and conclude the existence of a subsequence of $(\gmu)$
such that for $\mu\rightarrow 0$,
\begin{align*}
\gmu&\rightarrow g \mbox{ strongly in } \LDt\LD,\\
\gmu &\rightharpoonup g \mbox{ weakly in } \LDt{H^1(\D)},\\
\dt \gmu &\rightharpoonup \dt g \mbox{ weakly in } \LDt{\HDd}.
\end{align*}
Furthermore, by direct computation, we obtain
$$
\norm{\ct[g]g-\ct[\gmu]\gmu}{\LDt\LD}\leq \norm{\ct[g](g-\gmu)}{\LDt\LD}+\norm{(\ct[g]-\ct[\gmu])\gmu}{\LDt\LD}.
$$
The first term on the right side of the previous inequality goes to $0$ when $\mu\rightarrow 0$ because $\ct[\gmu]$ is bounded 
and $\gmu\rightarrow g $ strongly in $\LDt\LD$. Using Cauchy-Schwartz's inequality and that the domain $\Omega$ is bounded, yields
\begin{align*}
\norm{(\ct[g]-\ct[\gmu])\gmu}{L^1(0,T;L^1(\D))}=&\int_0^T \int_\D \bigg|\int_\D \big( \a h_1(\rj-\r)+\b \langle h_2(\rj-\r)\rangle\big) \times \\
& \times w(R-\Rj) \big(g(\rj,\Rj,t)-\gmu(\rj,\Rj,t)d\rj d\Rj\big)\bigg|\gmu(\r,R,t) d\r dR dt\\
 \leq& C \int_0^T\bigg(\int_{\D} g(\rj,\Rj,t)-\gmu(\rj,\Rj,t)d\rj d\Rj\bigg)\bigg(\int_{\D}\gmu(\r,R,t) d\r dR \bigg)dt\\
\leq& C |\D|^{\frac{1}{2}}\norm{\gmu-g}{\LDt\LD}^2.
\end{align*}
The constant is bounded from above by the $L^\infty$-norm of $h$ and
$w$, hence this term goes to $0$ as $\mu \rightarrow 0$.

Since $c[\gmu]\gmu $ is bounded, convergence holds in $L^p$ for all $p<\infty$.
The same argument holds for the difference $\norm{a[\gmu]\gmu-a[g]g}{\LDt\LD}$. So, we have shown that
\begin{align*}
& \ct[\gmu]\gmu \rightarrow \ct[g]g \mbox{ strongly in } \LDt\LD,\\
& a[\gmu]\gmu \rightarrow a[g]g \mbox{ strongly in } \LDt\LD. 
\end{align*}
Therefore, we can pass to the limit $\mu \rightarrow 0$ in the equation \eqref{eq17-w} and obtain for all $v \in \LDt{\HD}$
\be\label{eq27-w}
\int_0^T \Bigl\langle \dt g,v \Bigr\rangle \, dt= \int_0^T \int_{\D} a[g] g\deR v+\ct[g] g \dr v-\frac{\s}{2}\dr g \dr v \,dR d\r dt.
\ee
This completes the proof.
\end{proof}

\section{Long time behaviour of ratings and strength}\label{s:longtime}
\noindent In this section we study possible steady states of the proposed Elo
model and discuss the convergence of the ratings to the strength. We
recall that Junca and Jabin \cite{JJ2015} showed that the ratings of players converge to their intrinsic strength in the case $w=1$. This  corresponds to
the concentration of mass along the diagonal. In our model the
intrinsic strength is continuously increasing in time. Hence, to be
able to identify steady
states, we consider the shifted Fokker-Planck equation
\eqref{FP-scaling r-tilde}. 
Throughout this section we consider the problem in the whole space.

Since the diffusion part in \eqref{FP-scaling r-tilde} is singular,
the equation is degenerate parabolic. Degenerate Fokker-Planck
equations frequently, despite their lack of coercivity, exhibit
exponential convergence to equilibrium, a behaviour which has been
referred to by Villani as hypocoercivity in \cite{Villani2009}. For subsequent research on hypercoercity in linear Fokker-Planck equations, see \cite{arnoldErb,arnoldrevista}. Since \eqref{FP-scaling r-tilde} is a nonlinear, nonlocal Fokker-Planck equation these results do not apply here, but it is conceivable that generalisations of this approach can be used in studying the decay to equilibrium for \eqref{FP-scaling r-tilde}, which is however beyond the scope of the present paper. In the following, we present some results on the longterm behaviour of solutions to \eqref{FP-scaling r-tilde}.

Due to normalisation of the mean value, the only point
in which the formation of a steady state is possible are $R_0=0$ and $\r_0=0$. 
Let us assume that we have \revised{a} measure valued steady state in $(0,0)$, that is $g_{\infty}(\r,R)=\delta(\r)\delta(R)$. Then direct computations 
using the weak form of \eqref{FP-scaling r-tilde} give
\begin{align*}
0=&\dr (\phi(\r_0,R_0))[\a b(0)+\b \langle h_2\rangle(0)]+\sm w(0)\drr (\phi(\r_0,R_0))=\sm w(0)\drr (\phi(\r_0,R_0)).
\end{align*} 
This equation is not satisfied for all test functions $\phi$. Therefore, we investigate the possibility of having more complex steady states, 
which have a similar form as the one identified by Junca and Jabin. Let us assume that $g_{\infty}$ is of the form 
\be
g_{\infty}(\r,R)=\delta(\r)\tilde{g}(R),\label{delta rho}
\ee
or alternatively 
\be
g_{\infty}(\r,R)=\delta(R)\tilde{g}(\r)\label{delta R},
\ee
where $\tilde{g}(\cdot)$ in both cases is not a $\delta-$Dirac.\\
By direct computation in weak form of \eqref{FP-scaling r-tilde} with $\phi(\r,R)=\r^2$ and $\phi(\r,R)=R^2$ respectively, we compute the following expressions
for the second moments of the density function $g(\r,R,t)$: 
\begin{align}
\begin{split}
\frac{d}{dt}M_{g,2,\r}&(t)= \sm \int_{\R^4}w(R-\Rj)g(\r,R,t)g(\rj,\Rj,t)\,d \Rj d \ri dR d\r \\
&- \int_{\R^4}(\rj-\r)\big[\a b(\rj-\r)+\b \langle h_2(\rj-\r)\rangle\big]w(R-\Rj)g(\r,R,t)g(\rj,\Rj,t)\,d \Rj d \ri dR d\r , \label{2Momentscaling w}
\end{split}\\
\frac{d}{dt}M_{g,2,R}&(t)= \int_{\R^4}2R(b(\r-\rj)-b(R-\Rj))w(R-\Rj)g(\r,R,t)g(\rj,\Rj,t)\,d \Rj d \rj dR d\r.  \label{2-mom-R-g}
\end{align}
The analysis of the second moment w.r.t. $\r$ leads us to conclude that the diffusion prevents the formation of a steady state as in $\eqref{delta rho}$ if
$w=1$. Indeed, in this case, the first integral in \eqref{2Momentscaling w} equals $\s$. If at certain time  $\overline{t}>0$, $\r\simeq \rj$ or 
$g(\r,R,\overline{t})=\delta(\r-\r_0)\tilde{g}(R,\overline{t})$, the integral becomes small or vanishes 
(anyhow smaller than $\s$) and then $\frac{d}{dt}M_{2,\ri}(\overline{t})\geq 0$. Thus, we can conclude that the diffusion prevents the accumulation of the mass in $\r=0$. 
For a general choice of $w$, the long time behaviour of solutions is less clear.\\
Conversely, the second moment w.r.t. $R$ is decreasing. Due to the symmetry of the functions $b$ and $w$, we can rewrite $\eqref{2-mom-R-g}$ as
$$
\frac{d}{dt}M_{g,2,R}(t)=-\int_{\R^4}(R-\Rj)b(R-\Rj)w(R-\Rj)g(\r,R,t)g(\rj,\Rj,t)\,d \Rj d \rj dR d\r \leq 0.
$$
This inequality does not contradict the assumption of a steady state
of form  \eqref{delta R}. 

In order to evaluate if, with the scaling $\eqref{Scaling function}$, the rating converges to the intrinsic strength, let us define the energy
\begin{align}
E_2(t)&=\int_{\R^2} (\r-R)^2 g(\r,R,t) \,d\r dR.\label{energy2}
\end{align}
%For a general function $g(\r,R,t)$, it is not easy to evaluate the sign of the time-derivative of the energy $E_1(t)$, also in the case $w=1$. So, 
%we focus on the evaluating the evolution of the energy $E_2(t)$. \\
We are interested in the evolution of $E_2$ and compute 
\begin{align}
\begin{split}
\frac{d}{dt} E_2(t)=&-2\int_{\R^4}(\r-R) w(R-\Rj)b(R-\Rj)g(\r,R,t)g(\rj,\Rj,t)  \,d\rj d\Rj d \r dR\\
&+2\int_{\R^4}(\r-R)w(R-\Rj)b(\r-\rj)g(\r,R,t)g(\rj,\Rj,t) \, d\rj d\Rj d \r dR\\
&+2\a \int_{\R^4}(\r-R)w(R-\Rj)b(\r-\rj)g(\r,R,t)g(\rj,\Rj,t)\,  d\rj d\Rj d \r dR\\
&+2\b\int_{\R^4}(\r-R)w(R-\Rj) \langle h_2(\r-\rj)\rangle g(\r,R,t)g(\rj,\Rj,t)\,  d\rj d\Rj d \r dR\\
&+ \s\int_{\R^2}d[g]g(\r,R,t)\, d\r dR.\label{energy bis w}
\end{split}
\end{align} 
For general functions $w$ it is not possible to determine the signs of the respective integrals. Therefore, we consider the case $w=1$ only. 
For all odd functions $b(\cdot)$ (the same holds true for $\langle h_2(\r-\rj)\rangle$) we are able to show that
\begin{align*}
\int_{\R^4}\r b(\rj-\r) g(\r,R,t)&g(\rj,\Rj,t)  \,d\rj d\Rj d \r dR
\\&=\frac{1}{2}\int_{\R^4}\r (b(\rj-\r)- b(\r-\rj)) g(\r,R,t)g(\rj,\Rj,t)\,   d\rj d\Rj d \r dR\\
&=-\frac{1}{2} \int_{\R^4}(\rj-\r) b(\rj-\r)g(\r,R,t)g(\rj,\Rj,t) \, d\rj d\Rj d \r dR\\
&\le 0,
\end{align*}
and
$
\int_{\D^2}\r b(R-\Rj) g(\r,R,t)g(\rj,\Rj,t)\,  d\rj d\Rj d \r dR=0.
$
In this case we can rewrite the equation $\eqref{energy bis w}$ as
\begin{align}
\begin{split}
\frac{d}{dt} E_2(t)=-&\int_{\R^4}(R-\Rj)b(R-\Rj)g(\r,R,t)g(\rj,\Rj,t)  \,d\rj d\Rj d \r dR\\
&-\int_{\R^4}(\r-\rj)b(\r-\rj)g(\r,R,t)g(\rj,\Rj,t)  \,d\rj d\Rj d \r dR\\
&-\a \int_{\R^4}(\r-\rj)b(\r-\rj)g(\r,R,t)g(\rj,\Rj,t)  \,d\rj d\Rj d \r dR\\
&-2\b\int_{\R^4}(\r-\rj) \langle h_2(\r-\rj)\rangle g(\r,R,t)g(\rj,\Rj,t) \, d\rj d\Rj d \r dR\\
&+ \s.\label{energy bis}
\end{split}
\end{align}
Again we would like to know if a concentration of mass along the diagonal is possible. 
Let us assume that at certain time the solution is $g(\r,R,t)=\delta(\r-R)\tilde{g})(\r,R,t)$. \revised{If we} insert this claim in $\eqref{energy bis}$, we obtain
$$
\frac{d}{dt} E_2(t)=\s>0.
$$
It shows that the diffusion counteracts the accumulation of the mass
along the diagonal. On the other hand, the four integrals in $\eqref{energy bis}$ are strictly negative. Hence if $\s$ is small enough, the distance between rating 
and intrinsic strength becomes small, and the diffusive term can be
controlled. This indicates concentration of the mass in a certain neighbourhood of the diagonal in the long run.

\section{Numerical simulations}\label{s:numerics}

\noindent In this section we discuss the numerical discretisation of the Boltzmann equation \eqref{bolt-model-2-weight} and the shifted
Fokker-Planck equation \eqref{FP-scaling r-tilde}. We initialise the distribution of players with respect to their strength and 
rating with values from the unit interval and consider appropriately shifted interaction rules to ensure that the distribution
remains inside the unit square for all times $t > 0$.

\subsection{Monte Carlo simulations of the Boltzmann equation}

We use the classical Monte Carlo method to compute a series of realisations of the Boltzmann equation \eqref{bolt-model-2-weight}. In the direct Monte Carlo method, 
also known as Bird's scheme, pairs of players are randomly and non-exclusively selected for two-player games. The outcome of the game is determined by \eqref{e:interactionrules}. Note that
we consider the following shifted interaction rules for the ratings, to ensure that $\rho \in [0,1]$ and $R \in [0,1]$:
\begin{subequations}
 \begin{align}
\rit &= \ri + \var  \tilde{h} (\rj-\ri)w(\Ri-\Rj)+ \eta  \\
\rjt &= \rj + \var  \tilde{h} (\ri-\rj)w(\Ri-\Rj)+ \tilde{\eta},
 \end{align}
\end{subequations}
where $\tilde{h} = b(\rho_j - \rho_i)$. 
The microscopic interactions are simulated as follows: the outcome of the game $S_{ij}$ is the realisation of a discrete distribution function, which takes the value $\lbrace-1,1\rbrace$
with probability $\lbrace b(\rho_i-\rho_j),
1-b(\rho_i-\rho_j)\rbrace$.  The random variables $\eta$ are generated
such that they assume values $\eta = \pm 0.025$ with equal probability, the  parameter $\gamma$ is set to $0.05$.
Further information on Monte Carlo methods for Boltzmann type equations
can be found in \cite{PT}.

In each simulation we consider $N = 5000$ players and compute the steady state distribution by performing $10^8$ time steps. 
The result is then averaged over another $10^5$ time steps. We perform
$M=10$ realizations and compute the density from the averaged steady
states. 

\subsection{Finite volume discretisation and simulations of the nonlinear Fokker-Planck equation}

The solver for the Fokker-Planck equation is based on a Strang splitting and an upwind finite volume scheme. 
We recall that we discretise the shifted Fokker-Planck equation \eqref{FP-scaling r-tilde}, which allows us to perform simulations on a bounded domain. 
Because of the splitting we consider the interactions in the rating and the strength variable separately. We define two operators, which correspond to
\begin{enumerate}[label = ($\mathcal{S}_\arabic*$):]
\item Interaction step in the strength variable $R$:
\begin{align*}
\frac{\partial g^*}{\partial t}(\rho, R, t) = -\frac{\partial}{\partial \rho} (c[\tilde{g}] g^*(\rho, R, t)) + \frac{\sigma^2}{2} d[\tilde{g}]\frac{\partial^2}{\partial \rho^2} ( g^*(\rho, R, t))
\end{align*}
subject to the initial condition $g^*(\rho,R,t) = \tilde{g}(\rho,R,t)$. Note that we compute the interaction integrals using $\tilde{g}$, which corresponds to the solution at the previous
time step in the full splitting scheme. 
\item Interaction step in the rating variable $\rho$:
\begin{align*}
\frac{\partial g^\diamond}{\partial t}(\rho, R, t) = -\frac{\partial}{\partial R} (a[g^*] g^\diamond(\rho, R, t))
\end{align*}
\end{enumerate}
We approximate all integrals, which appear in the interaction
coefficients using the trapezoidal rule.

Let $\hat{g}^k$ denote the solution at time $t^k = k \Delta t$, where $\Delta t$ corresponds to the time step size.  Then the Strang splitting results in the scheme
\begin{align*}
\hat{g}^{k+1}(\rho,R) = \mathcal{S}_2\Bigl(\hat{g}^{*,k+1}, \frac{\Delta t}{2}\Bigr) \circ \mathcal{S}_1\Bigl(\hat{g}^{\diamond,k+\frac{1}{2}},\Delta t\Bigr) \circ \mathcal{S}_2\Bigl(\hat{g}^k,\frac{\Delta t}{2}\Bigr),
\end{align*}
where the superscripts denote the solutions of $g^*$ and $g^\diamond$ at the discrete time steps $t^{k+1} = (k+1) \Delta t$ and $t^{k+\frac{1}{2}} = (k + \frac{1}{2}) \Delta t$. 
We use a conservative upwind finite volume discretisation to discretise the respective operators. The corresponding explicit-in-time upwind finite volume methods is given by
\begin{align*}
\hat{g}_j^{n+1} = \hat{g}_j^n + \lambda_1 (\hat{c}_{j + \frac{1}{2}} - \hat{c}_{j - \frac{1}{2}})  + \lambda_2(\hat{d}_{j + \frac{1}{2}} - \hat{d}_{j - \frac{1}{2}}),
\end{align*}
where $\hat{c}$ is the upwind flux and the diffusive flux is given $\hat{d}_{j+\frac{1}{2}} = D(\hat{g}_{j+1}) \hat{g}_{j+1} - D(\hat{g}_{j}) \hat{g}_j$. Here $\lambda_1 = {\Delta t}/{\Delta x}$ and $\lambda_2 ={\Delta t}/{\Delta x^2}$.

\subsection{Computational experiments}

All micro- and macroscopic simulations are performed on the domain
$[0,1]\times[0,1]$ with no-flux boundary conditions.  In the case of a
general interaction function, the interaction rate function $w(r_i-r_j)$ is a piecewise constant function given by
\begin{align}
w(z) = 
\begin{cases}
1 \quad \text{ if } \lvert z \rvert \leq 0.1\\
0 \quad \text{ otherwise.}
\end{cases}
\label{e:compactw}
\end{align}

\subsubsection{All-play-all tournaments:} 
\begin{figure}[htb!]
\begin{center}
\subfigure[Steady state  -- no diffusion]{\includegraphics[width=0.4\textwidth]{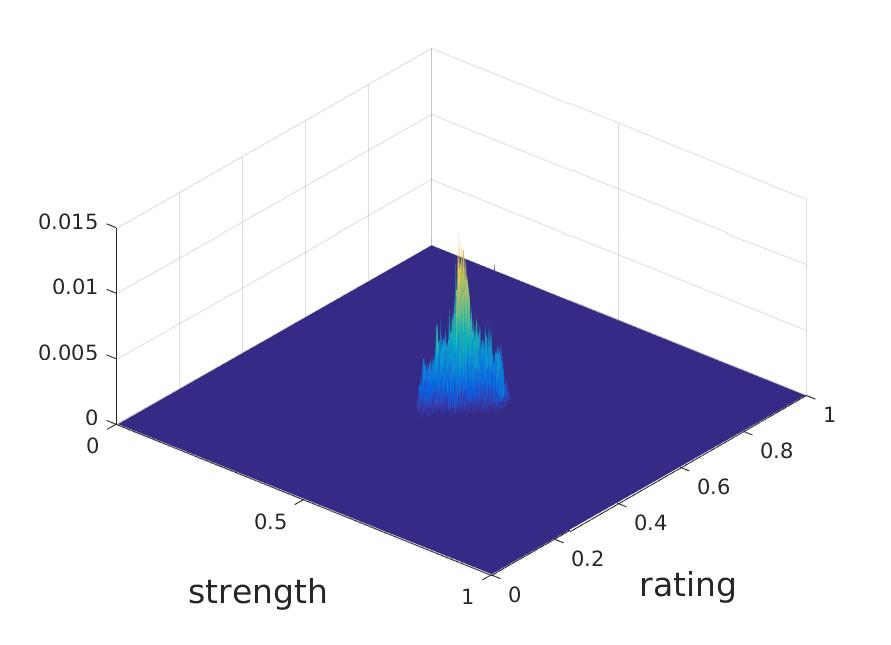}}\hspace*{1cm}
\subfigure[Steady state  (top view) -- no diffusion]{\includegraphics[width=0.4\textwidth]{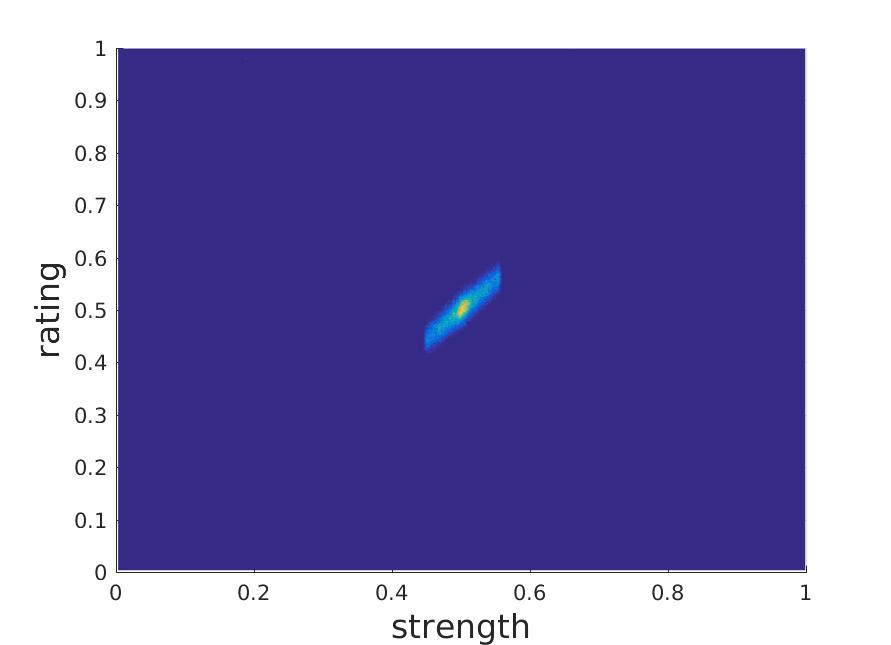}}\\
\subfigure[Steady state -- diffusion $\nu= 0.025 $]{\includegraphics[width=0.4\textwidth]{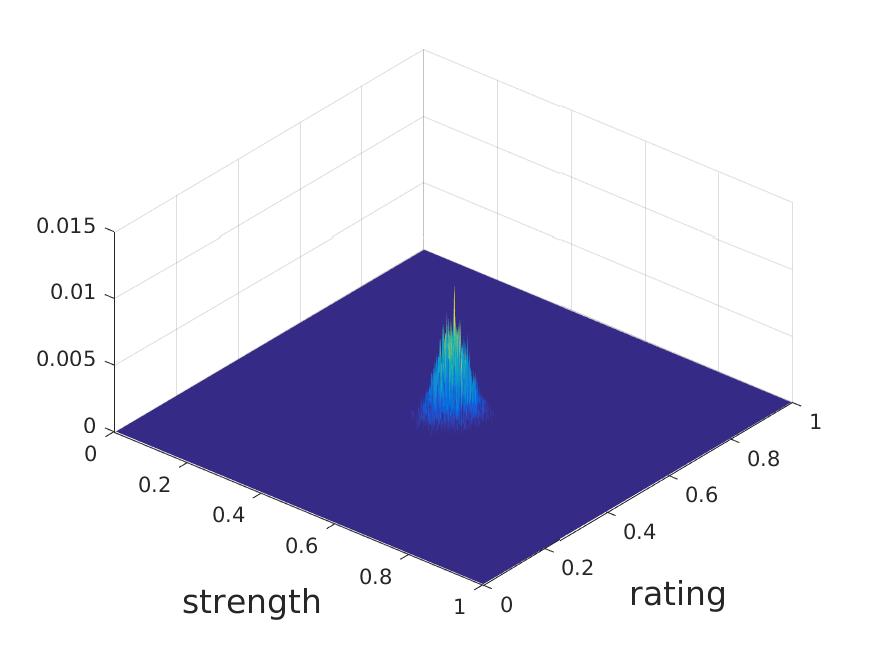}}\hspace*{1cm}
\subfigure[Steady state  (top view) -- diffusion $\nu = 0.025$]{\includegraphics[width=0.4\textwidth]{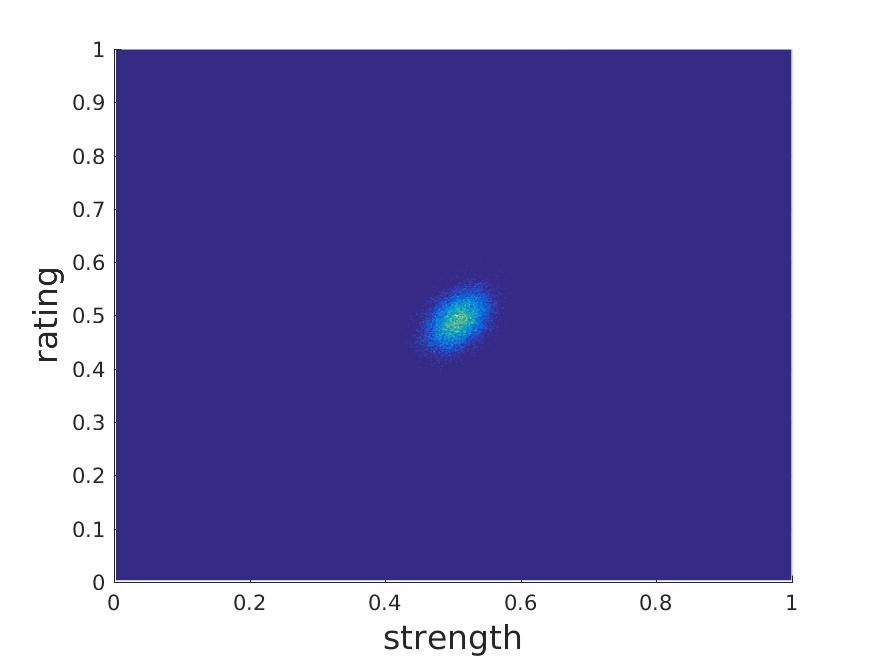}}
\caption{Computational steady state of the Boltzmann model for $w=1$ in the case of no diffusion, $\eta = \bar{\eta}= 0$, and small diffusion in the strength $\eta = \bar{\eta} = 0.025$.}
\label{fig:gen_elo_w1_micro}
\end{center}
\end{figure}

\begin{figure}[htb!]
\begin{center}
\subfigure[Steady state (top view) -- no diffusion]{\includegraphics[width=0.4\textwidth]{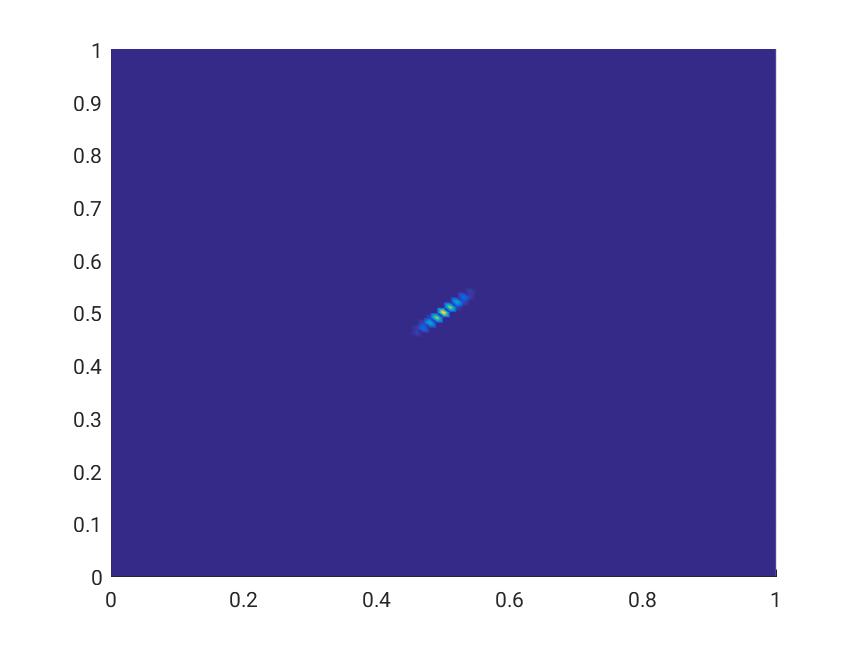}}\hspace*{1cm}
\subfigure[Steady state (top view) -- with diffusion]{\includegraphics[width=0.4\textwidth]{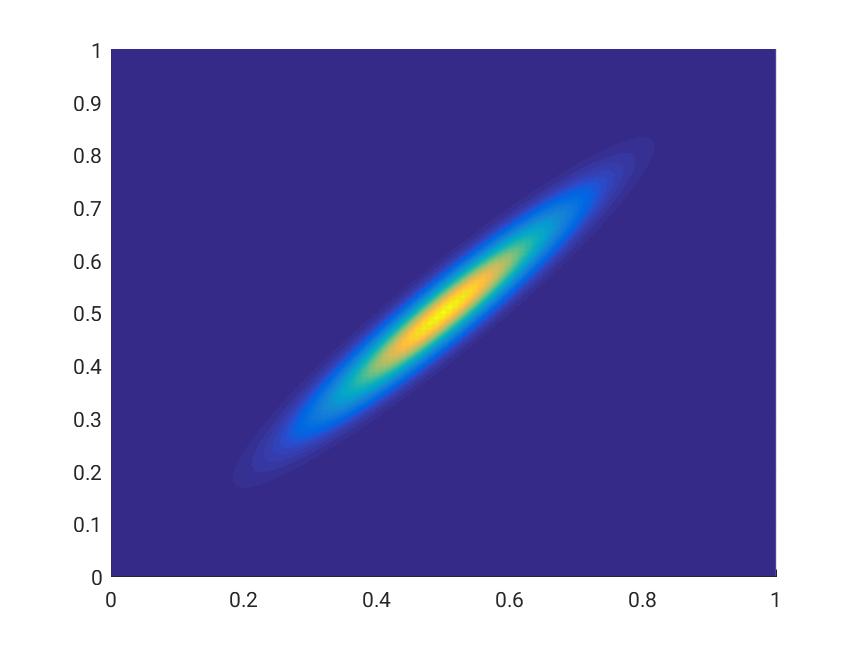}}\\
\subfigure[Energy decay in the case of no diffusion]{\includegraphics[width=0.4\textwidth]{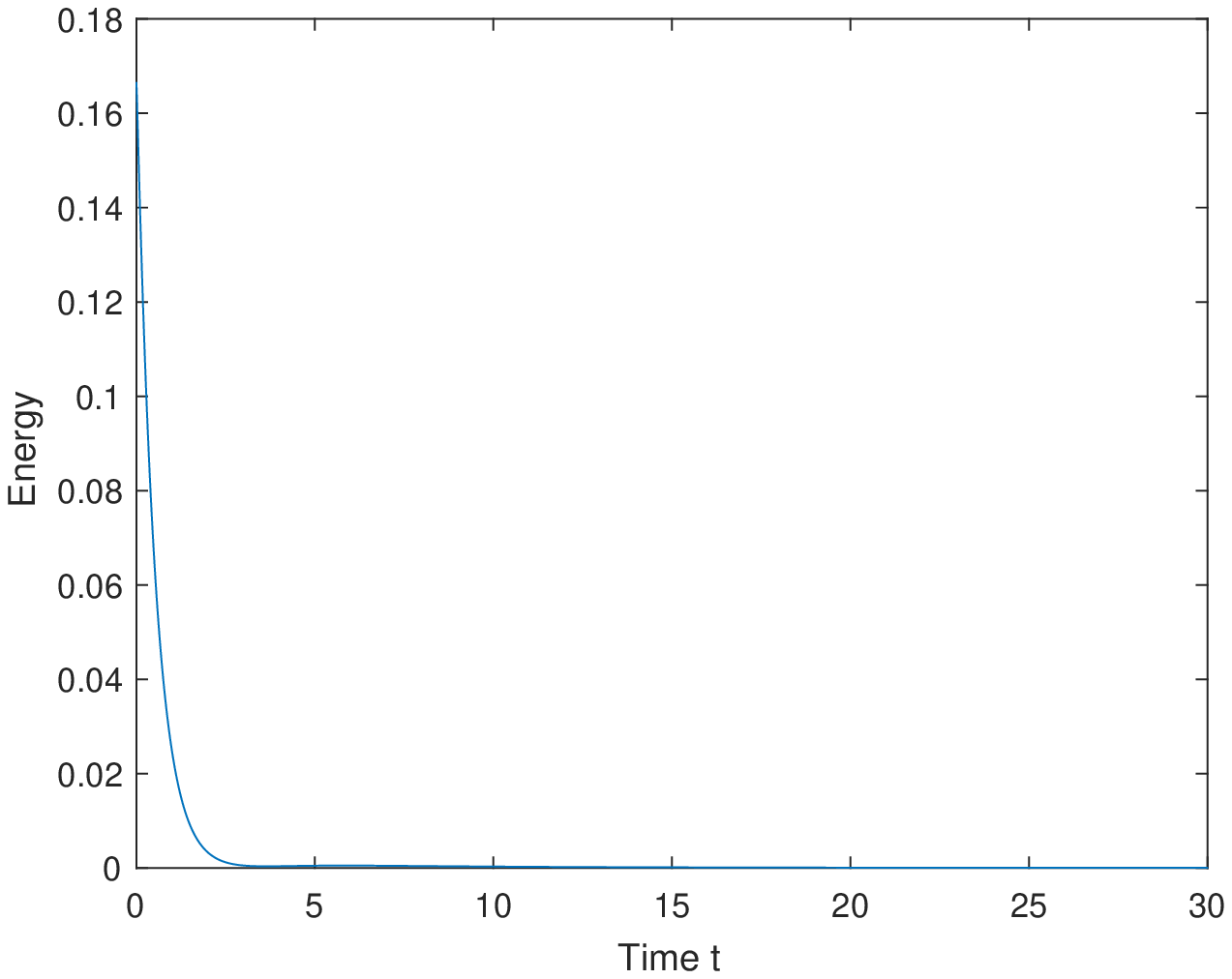}} \hspace*{1em}
\subfigure[Energy decay in the case of diffusion]{\includegraphics[width=0.4\textwidth]{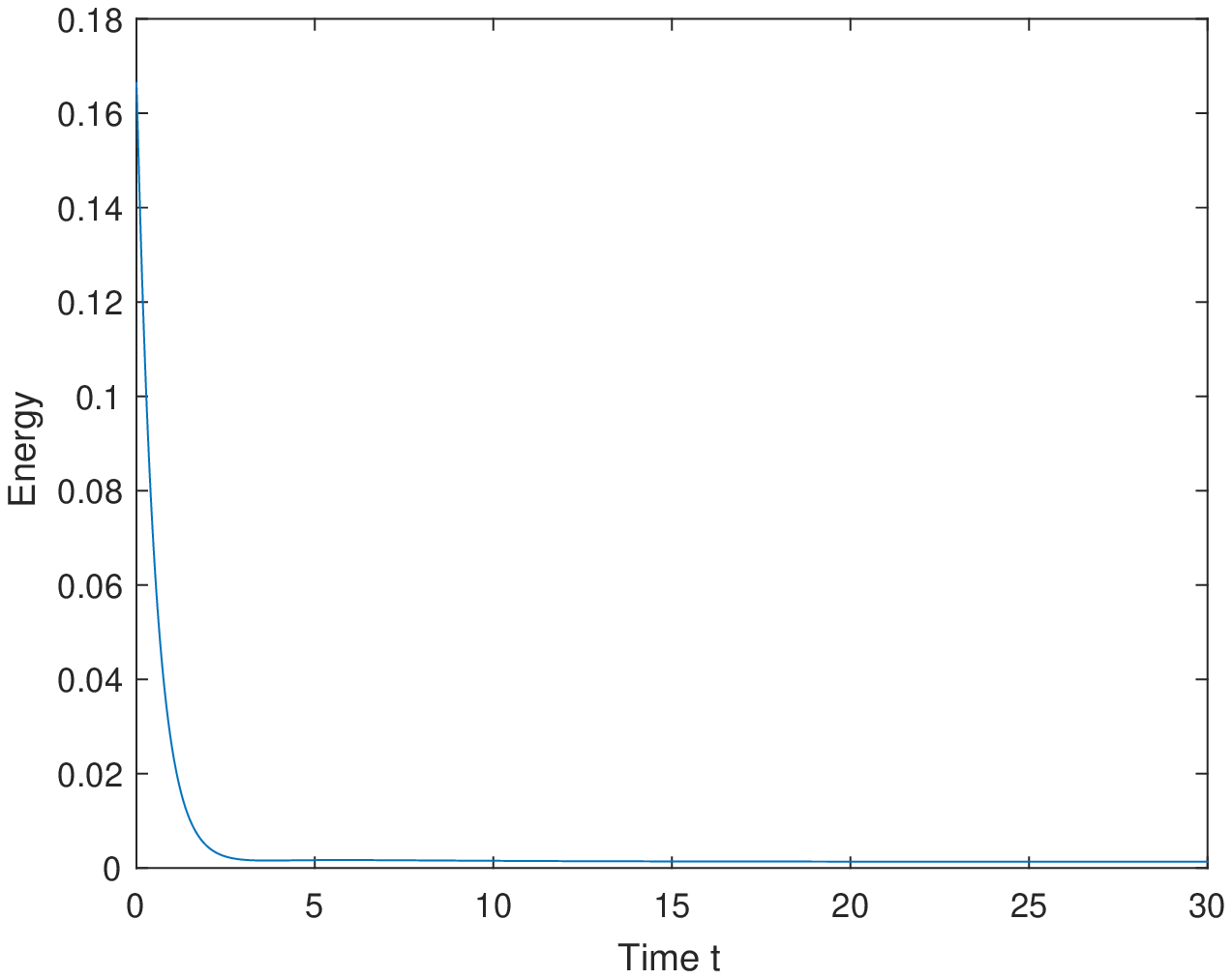}}
\caption{Computational steady state of the Fokker-Planck model and
  energy decay for
  $w=1$ in the case of no and little diffusion strength.}\label{fig:gen_elo_w1_macro}
\end{center}
\end{figure}

We start by investigating the long time behaviour of the Elo model with $w=1$, $\alpha=0.1$ and $\beta=0$ in \eqref{e:h}. 
Hence players have the same probability to play
against another independent of their respective ratings. We have seen in Section \ref{s:longtime} that we expect a measure valued solution in the case of no diffusion.
However, we can not show convergence of solutions to a measure valued
steady state if stochastic fluctuations influence the intrinsic strength. 
In the following we compare computed steady states of the Boltzmann as well as the Fokker-Planck equation in the case of diffusion and no diffusion.
We start with a uniform distribution
of agents in the micro- as well as the macroscopic situation.  Figure~\ref{fig:gen_elo_w1_micro} as well as Figure~\ref{fig:gen_elo_w1_macro} confirm the expected
formation of a Delta Dirac at the center of mass in the case of no diffusion. If the individual strength is also influenced by stochastic fluctuations, the 
steady state is smoothed out with respect to the rating as well. The resulting steady states are Gaussian like profiles in the micro- as well as macroscopic 
simulations, see  Figures \ref{fig:gen_elo_w1_micro} and \ref{fig:gen_elo_w1_macro}. Figure \ref{fig:gen_elo_w1_macro} also shows the decay of the energy $E_2$ in time. 

\subsubsection{\revised{Competitions of players with similar ratings}}
\noindent Assigning initial ratings to players in the Elo rating is a delicate
issue, since inaccurate initial ratings may 
influence the ability of the rating to converge to a `good' rating of
players reflecting their intrinsic strengths. We \revised{show the difficulties in this case} by studying \revised{the dynamics if players with close ratings compete}. 

We set the interaction rate function
to \eqref{e:compactw} -- hence individuals only play against each other, if the difference between their ratings is
small. We consider two groups of players with different strength and rating levels as initial distribution. The first group is underrated, that is all players have rating $R = 0.2$ but their strength is 
distributed as $\rho \in \mathcal{N}(0.75, 0.1)$. The second group is
overrated, with rating $R=0.9$ and a uniform distribution in
strength. 
We use this initial configuration in two computational experiments.

\begin{figure}[htb!]
\begin{center}
\subfigure{\includegraphics[width=0.4\textwidth]{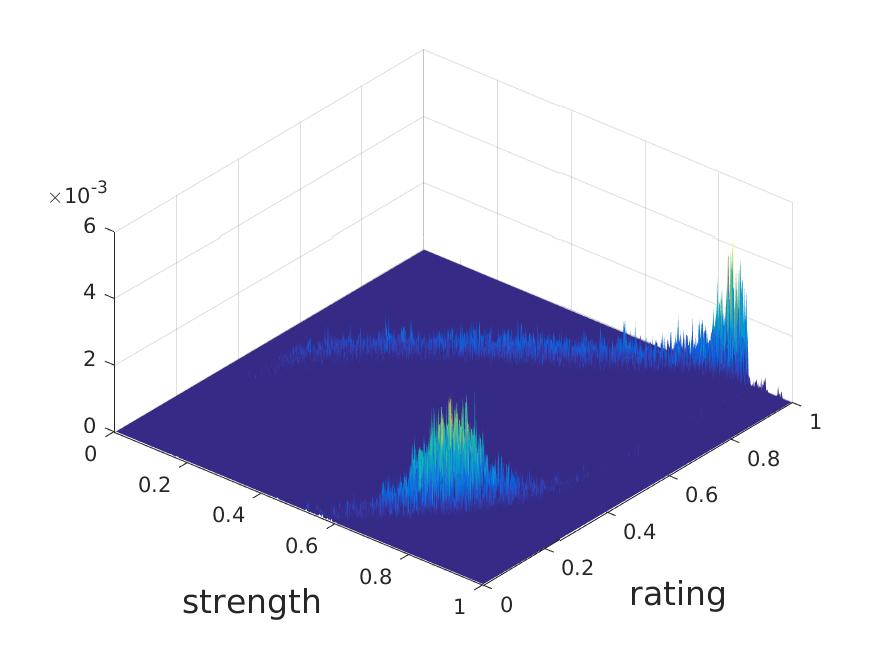}}\hspace*{1cm}
\subfigure{\includegraphics[width=0.4\textwidth]{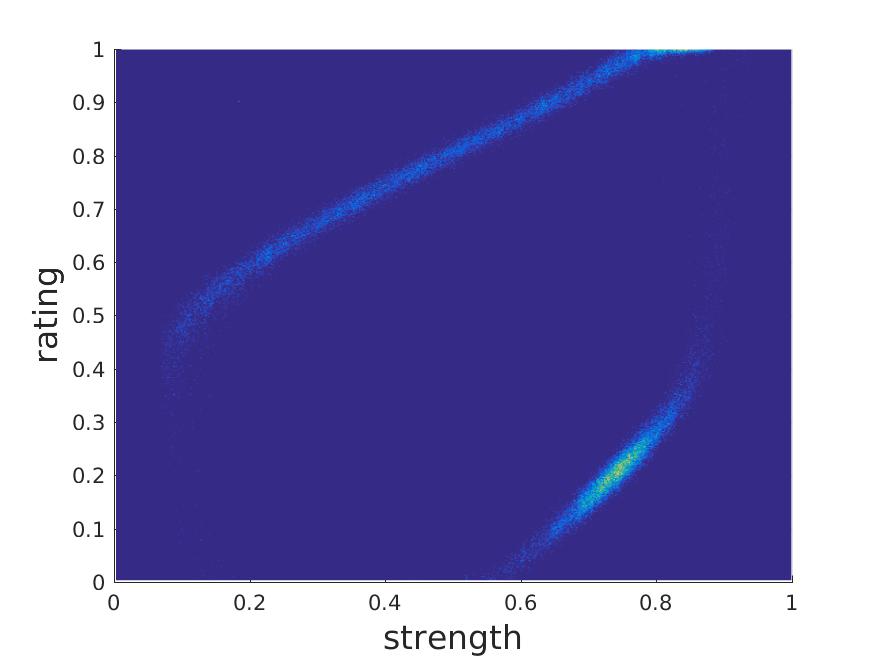}}
\caption{Computed stationary profiles in \revised{competitions of players with similar ratings} in case of two initially separated groups (one underrated with  high strength but low rating and one overrated
 with variable strength but rating $0.9$). Due to the limited interaction between the groups and the chosen learning mechanism, they remain separated.}\label{f:twogroups}
\end{center}
\end{figure}

In the first, we choose the learning parameters $\alpha=0.1$ and $\beta=0$. We see that the two groups remain separated due to their different ratings in this case, see Figure \ref{f:twogroups}. However, players compete within
their own group and since $\beta=0$ the overall rating improves. In the overrated group the strongest players accumulate at the highest possible rating, while the underrated
group forms a diagonal pattern. Here the underrated players evolve to
the maximum possible rating level.

\begin{figure}[htb!]
\begin{center}
\subfigure{\includegraphics[width=0.4\textwidth]{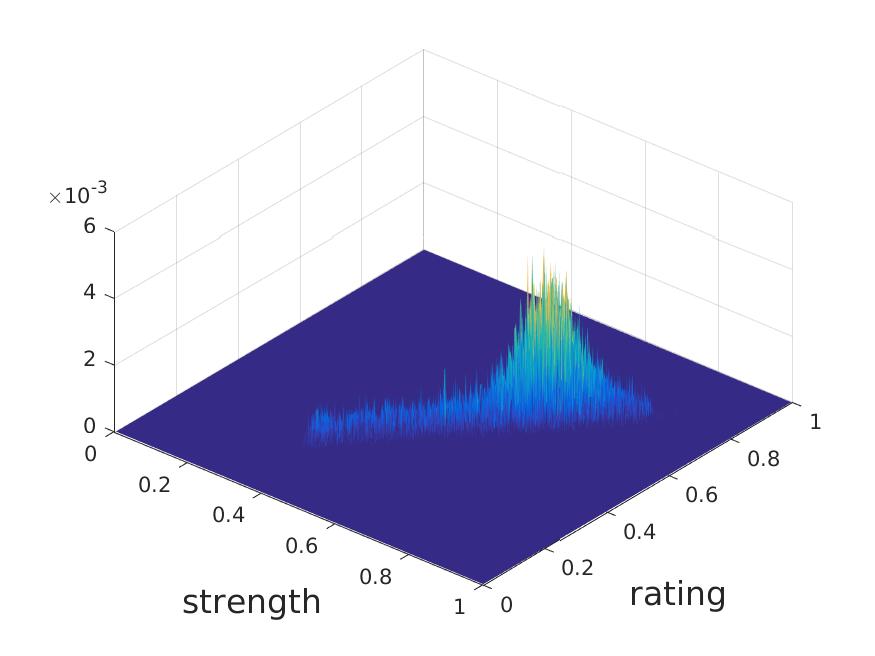}}\hspace*{1cm}
\subfigure{\includegraphics[width=0.4\textwidth]{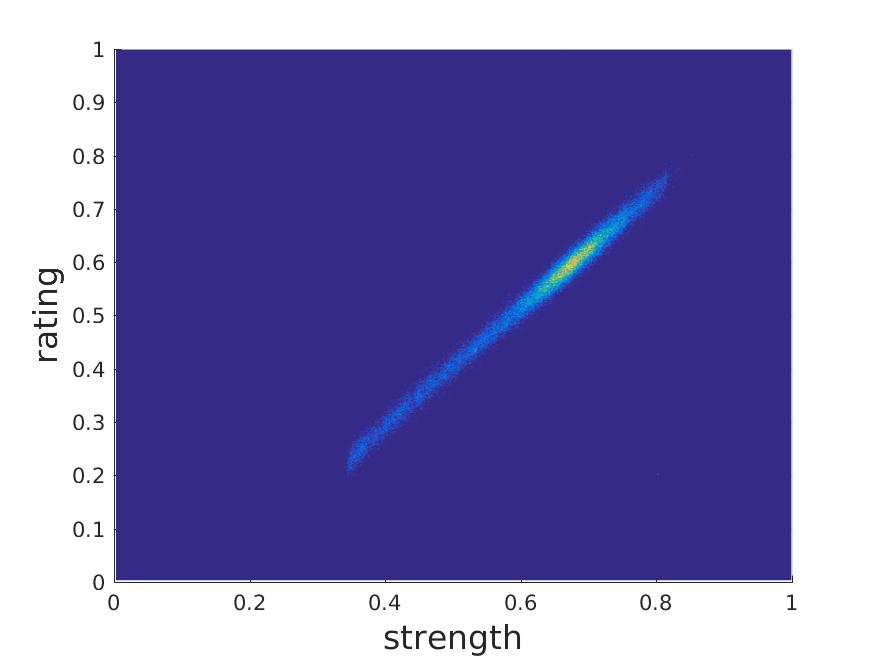}}
\caption{Computed stationary profiles in \revised{competitions of players with similar ratings} in case of two initially separated groups (one underrated with  high strength but low rating and one overrated
 with variable strength but rating $0.9$). Despite the limited
 interaction between the groups, the adapted learning mechanism leads
 to convergence of the ratings to a slightly shifted diagonal.}\label{f:twogroups2}
\end{center}
\end{figure}

In the second experiment, using the same initial configuration, but $\alpha=0.1$ and $\beta = 0.05$ the
steady state profile looks totally different. In this setting stronger players loose strength, when loosing against a weaker
opponent. Therefore, the ratings of the overrated group decrease, while the ratings of the underrated group increases. After a while the two groups merge, accumulating on a diagonal
which underestimates the intrinsic strength of players by
approximately $0.1$, see Figure \ref{f:twogroups2}.

These examples show the importance of the initial ratings as well as the
influence of the adapted learning mechanism.

\subsubsection{Foul play}
\begin{figure}
 \includegraphics[width=0.4\textwidth]{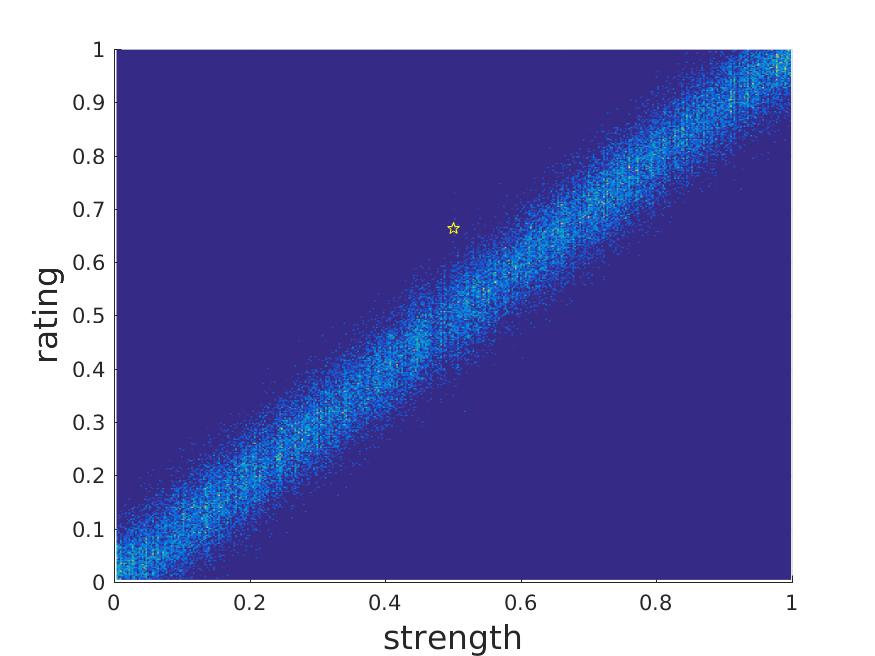}
 \caption{Computed stationary profile in a foul play where the first
   player has an unfair advantage in each game. We observe that the ratings and strength all players except the first one converge. 
 The cheating player (indicated by a star) ends up with a higher rating than it is supposed to have.}\label{f:cheater}
\end{figure}

Finally, we consider a series of games, in which one player, without
loss of generality the first one, is playing unfairly, e.g.\ through
cheating, doping or bribing of referees. This means that the outcome of every \revised{microscopic} game which involves this player
is biased in their favor. In particular we assume that the probability of winning is increased by a factor $\tilde{b}$ for player 1 and decreased by $\tilde{b}$ for the other
contestant. Figure~\ref{f:cheater} shows the stationary profile in the case of a uniform initial distribution of agents, $\alpha = 0.1$, $\beta = 0$, $w=1$ and $\tilde{b} = 0.2$.
The star indicates  the position of the unfair first player. While the distribution of players with respect to their ratings and their strengths accumulates along the diagonal,
we see that the first player is rated higher than implied by
\revised{his or her} strength. 

\appendix
\section{Derivation of the Fokker-Planck equation}\label{s:appendix}
\noindent In this section we derive the limiting Fokker-Planck
equation in the case $\gamma \rightarrow 0$, $\sigma_{\eta}
\rightarrow 0$ such that ${\sigma_{\eta}^2}/{\gamma} =: \sigma^2$ is
kept fixed.
Based on the interaction rules \eqref{e:interactionrules}, which define the outcome of a game, we compute the expected values of the following quantities:
\begin{align*}
\langle(\Rit-\Ri)\rangle &=\var (b(\ri-\rj)-b(R_i-R_j))\\
\langle((\Rit-\Ri)^2)\rangle &=\var ^2(b(\ri-\rj)-b(R_i-R_j))^2;\\
\langle(\rit-\ri)\rangle &= \var (\a h_1(\rj-\ri)+\b \langle h_2(\rj-\ri)\rangle )\\
\langle(\rit-\ri)^2\rangle &=\var^2 \left(\a h_1(\rj-\ri)+\b \langle h_2(\rj-\ri)\rangle\right)^2+\sigma_{\eta} ^2 \\
\langle(\rit-\ri)(\Rit-\Ri)\rangle &=\var^2 (\a h_1(\rj-\ri)+\b \langle h_2(\rj-\ri)\rangle)(b(\ri-\rj)-b(\Ri-\Rj)).
\end{align*}
Using Taylor expansion of $\phi(\rit,\Rit)$ up to order two around $(\ri,\Ri)$, we obtain
\begin{align*}
\langle &\phi(\rit,\Rit)-\phi(\ri,\Ri) \rangle \\
&= \langle \Rit-\Ri \rangle\deRi \phi(\ri,\Ri)+ \langle \rit-\ri \rangle\deri \phi(\ri,\Rj)\\
&+\frac{1}{2}\bigg[\langle(\Rit-\Ri)^2\rangle\deRiRi \phi(\ri,\Ri)+ \langle(\rit-\ri)^2\rangle \deriri \phi(\ri,\Ri)+2 \langle(\rit-\ri)(\Rit-\Ri)\rangle \frac{\partial^2}{\partial \ri \partial \Ri}\phi(\ri,\Ri)\bigg]\\
&+\mathcal{R}_\var(\phi,\rit,\Rit,\ri,\Ri,\t), 
\end{align*}
where the remainder term $\mathcal{R}_{\var}$ is given by
$$\mathcal{R}_\var=\begin{pmatrix}
\rit-\ri \\ \Rit-\Ri
\end{pmatrix}^T
\begin{pmatrix}
\deriri \phi(\overline{\ri},\overline{\Ri})-\deriri \phi(\ri,\Ri) & \frac{\partial^2}{\partial \ri \partial \Ri}\phi(\overline{\ri},\overline{\Ri})-\frac{\partial^2}{\partial \ri \partial \Ri}\phi(\ri,\Ri)\\
\frac{\partial^2}{\partial \ri \partial \Ri}\phi(\overline{\ri},\overline{\Ri})-\frac{\partial^2}{\partial \ri \partial \Ri}\phi(\ri,\Ri) & \deRiRi \phi(\overline{\ri},\overline{\Ri})-\deRiRi \phi(\ri,\Ri)
\end{pmatrix}
\begin{pmatrix}
\rit-\ri \\ \Rit-\Ri
\end{pmatrix},
$$
for some $0\leq \theta_1, \theta_2\leq 1$ with $\overline{\ri}$ and $\overline{\Ri}$  defined as 
 $$\overline{\ri}= \theta_1 \ri+(1-\theta_1)\rit \text{ and } \overline{\Ri}= \theta_2 \Ri+(1-\theta_2)\Rit.$$
Next we rescale time as $\t=\var t$ and insert the expansion in \eqref{bolt-model-2-weight}. This yields 
\be 
\begin{split}
&\frac{d}{d\t}\int_{\R^2} \phi(\ri,\Rj)\fe(\ri,\Ri,\t) d\Ri d\ri= \frac{1}{2\var}\int_{\R^2}\tilde{\mathcal{R}}_\var(\phi,\rit,\Rit,\ri,\Ri,\t)\fe(\ri,\Ri,\t)d\Ri d\ri\\
&+\int_{\R^4} \bigg[\deRi \phi(\ri,\Rj)(b(\ri-\rj)-b(R_i-R_j))
+\deri \phi(\ri,\Rj)(\a h_1(\rj-\ri)+\b \langle h_2(\rj-\ri)\rangle)\\
&\qquad +\frac{\sigma_{\eta}^2}{2\gamma}\deriri \phi(\ri,\Rj) \bigg] w(\Ri-\Rj)\fe(\ri,\Ri,\t) \fe(\rj,\Rj,\t)  d \Rj d\rj d \Ri d\ri ,
\end{split}
\ee
where 
\begin{align*}
\begin{split}
\tilde{\mathcal{R}}_\var&(\phi,\rit,\Rit,\ri,\Ri,\t)=\var^2\int_{\R^2} \deRiRi \phi (\ri,\Ri) (b(\ri-\rj)-b(\Ri-\Rj))^2w(\Ri-\Rj)\fe(\rj,\Rj,\t)d\Rj d\rj\\
&+\var^2\int_{\R^2}  \deriri \phi (\ri,\Ri)\big( \a h_1(\rj-\ri)+\b \langle h_2(\rj-\ri)\rangle\big)^2w(\Ri-\Rj)\fe(\rj,\Rj,\t)d\Rj d\rj\\
&+2\var^2 \int_{\R^2} \frac{\partial}{\partial \ri \partial\Ri} \phi (\ri,\Ri) (b(\ri-\rj)-b(\Ri-\Rj))\big( \a h_1(\rj-\ri)+\b \langle h_2(\rj-\ri)\rangle\big)  w(\Ri-\Rj)\fe(\rj,\Rj,\t)d\Rj d\rj\\
&+\int_{\R^2}R_\var w(\Ri-\Rj)\fe(\rj,\Rj,\t)d\Rj d\rj.
\end{split}
\end{align*}
Next we show that the remainder $\frac{1}{2\var}\int_{\R^2}\tilde{\mathcal{R}}_\var(\phi,\rit,\Rit,\ri,\Ri,\t)\fe(\ri,\Ri,\t)d\Ri d\ri$  vanishes for $\var\rightarrow 0$.
Let us assume that $\phi(\ri,\Ri)$ belongs to the space $\mathcal{C}_{2+\delta}(\R^2)=\{h:\R^2\rightarrow\R, \ \|D^\zeta h\|_\delta<+\infty\}$, where $0<\delta\leq 1$, $\zeta$ is a multi-index with $|\zeta|\leq2$ and the seminorm $\|\cdot\|_\delta$ is the usual H{\"o}lder seminorm
 $$
 \|f\|_\delta=\sup_{x,y\in \R^2}\frac{|f(x)-f(y)|}{|x-y|^\delta}.
 $$
With this choice of $\phi(\ri,\Ri)$, all the terms wich contain $\deriri \phi$ and $\deRiRi \phi$ vanish using the same arguments as in \cite{T,CPP}. Hence, 
we focus on the mixed derivative $\frac{\partial^2}{\partial \ri \partial \Ri}\phi(\ri,\Ri)$. Since $\phi(\ri,\Ri)\in  \mathcal{C}_{2+\delta}(\R^2)$ and 
$\|(\overline{\ri},\overline{\Ri})-(\ri,\Ri)\|\leq\|(\rit,\Rit)-(\ri,\Ri)\|$, we  have
$$
\left|\frac{\partial^2}{\partial \ri \partial \Ri}\phi(\overline{\ri},\overline{\Ri})-\frac{\partial^2}{\partial \ri \partial \Ri}\phi(\ri,\Ri) \right|
\leq \|\phi\|_{2+\delta}\|(\rit,\Rit)-(\ri,\Ri)\|^\delta.
$$
Furthermore, due to \eqref{function b}, \eqref{h1} and \eqref{h2odd},
$$
\|(\rit,\Rit)-(\ri,\Ri)\|=\left[\var^2\left( \a h_1(\rj-\ri)+\b \langle h_2(\rj-\ri)\rangle\right)^2 +\var^2 \left(b(\ri-\rj)-b(\Ri-\Rj) \right)^2\right]^{\frac{1}{2}}\leq C \var.
$$
Using the previous inequalities we estimate the mixed term as
\begin{align*}
&\frac{1}{2\var}\left|\int_{\R^4} \left(\frac{\partial^2\phi(\overline{\ri},\overline{\Ri})}{\partial \ri \partial \Ri}-\frac{\partial^2\phi(\ri,\Ri)}{\partial \ri \partial \Ri}\right)
 \left\|\begin{pmatrix}
\overline{\ri}\\\overline{\Ri}
\end{pmatrix}-\begin{pmatrix}\ri \\ \Ri\end{pmatrix}\right\|^2
  w(\Ri-\Rj)\fe(\rj,\Rj,\t)\fe(\ri,\Ri,\t)\, d\Ri d\ri d\Rj d\rj \right|\\
&\phantom{\frac{1}{2\var}\int_{\R^4}}\leq\frac{1}{2\var}\int_{\R^4}\|\phi\|_{2+\delta}\|(\rit,\Rit)-(\ri,\Ri)\|^\delta\|(\rit,\Rit)-(\ri,\Ri)\|^2\fe(\rj,\Rj,\t)\fe(\ri,\Ri,\t)\,d\Ri d\ri d\Rj d\rj\\
&\phantom{\frac{1}{2\var}\int_{\R^4}}\leq \frac{1}{2\var}\int_{\R^4}C^\delta\|\phi\|_{2+\delta}\var^{2+\delta}\fe(\rj,\Rj,\t)\fe(\ri,\Ri,\t)\,d\Ri d\ri d\Rj d\rj\leq \frac{C^\delta}{2} \|\phi\|_{2+\delta} \var^{1+\delta}.
\end{align*}
Hence the remainder term converges to $0$ as $\var\rightarrow 0$. Therefore, the density $\fe(\ri,\Ri,\t)$ converges to $f(\ri,\Ri,\t)$ 
which solves
\begin{align} 
\begin{split}
\frac{d}{d\tau}&\int_{\R^2} \phi(\ri,\Rj)f(\ri,\Ri,\t) d\Ri d\ri= \\
& \int_{\R^2}f(\ri,\Ri,\t)\Bigg\{
\deRi \phi(\ri,\Rj)\bigg[ \int_{\R^2} w(\Ri-\Rj)(b(\ri-\rj)-b(\Ri-\Rj))f(\rj,\Rj,\t)d\rj d\Rj\bigg]\\
&+\deri\phi(\ri,\Rj) \bigg[\int_{\R^2}w(\Ri-\Rj)\big( \a h_1(\rj-\ri)+\b \langle h_2(\rj-\ri)\rangle\big) f(\rj,\Rj,\t)d\rj d\Rj \bigg]  \\
&+\frac{\s}{2} \deriri \phi(\ri,\Rj)  \bigg[  \int_{\R^2}w(\Ri-\Rj) f(\rj,\Rj,\t)d\rj d\Rj\bigg]\Bigg\}d\Ri d\ri.\label{weak-FP-model 2}
\end{split}
\end{align}
It remains to show that under suitable boundary conditions equation \eqref{weak-FP-model 2} gives the desired weak formulation of the Fokker Planck equation. 
We split the boundary terms $BT$ into the different parts $BT_i$, $i=1,2,3$ that arises respectively from each integral. They are given by
\begin{align*}
B_1=&\inr \left[f (\ri,\Ri,\t) \phi (\ri,\Ri)\left(\int_{\R^2} w(\Ri-\Rj)(b(\ri-\rj)-b(R_i-R_j))f(\rj,\Rj,\t) d\Rj d\rj\right)\right]_{\Ri=-\infty}^{\Ri=+\infty} d\ri\\
B_2=&\inr \bigg[f (\ri,\Ri,\t) \phi (\ri,\Ri)\left(\int_{\R^2} w(\Ri-\Rj)(\a h_1(\rj-\ri)+\b \langle h_2(\rj-\ri)\rangle)f(\rj,\Rj,\t) d\Rj d\rj\right)\bigg]_{\ri=-\infty}^{\ri=+\infty} d\Ri\\
B_3=&\frac{\s}{2}\inr \bigg[  \deri \phi (\ri,\Ri) f (\ri,\Ri,\t)  \bigg( \int_{\R^2}w(\Ri-\Rj) f(\rj,\Rj,\t)d\rj d\Rj	\bigg) \\
& -\phi (\ri,\Ri) \deri \bigg[f (\ri,\Ri,\t)  \bigg( \int_{\R^2}w(\Ri-\Rj) f(\rj,\Rj,\t)d\rj d\Rj	\bigg) \bigg]    \bigg]_{\ri=-\infty}^{\ri=+\infty} d\Ri.
\end{align*}
These three terms are zero, if the following boundary conditions are satisfied: 
\begin{align}
&\lim_{|\Ri|\rightarrow +\infty} f(\ri,\Ri,\t)=0 , ~~\lim_{|\ri|\rightarrow +\infty} f(\ri,\Ri,\t)=0,~\lim_{|\ri|\rightarrow +\infty} \deri f(\ri,\Ri,\t)=0.\label{bc}
\end{align}
These boundary condition are guaranteed for the Boltzmann equation $\fe(\ri,\Ri,\t)$ by mass conservation and the upper and lower bounds on the mean, see \eqref{1-mom-bolt}. 
Therefore, \eqref{weak-FP-model 2} is the weak form of the Fokker-Planck equation
\begin{align} 
\begin{split}
\frac{d}{d\tau}&\int_{\R^2} \phi(\ri,\Ri)f(\ri,\Ri,\t) d\Ri d\ri=\\
&\int_{\R^2}\phi(\ri,\Ri)\bigg\{-\deRi\bigg[f(\ri,\Ri,\t) \int_{\R^2} w(\Ri-\Rj)(b(\ri-\rj)-b(\Ri-\Rj))f(\rj,\Rj)d\rj d\Rj\bigg]\\
&-\deri \bigg[ f(\ri,\Ri,\t) \int_{\R^2}w(\Ri-\Rj)\big( \a h_1(\rj-\ri)+\b \langle h_2(\rj-\ri)\rangle\big)f(\rj,\Rj)d\rj d\Rj\bigg] \\
&+\frac{\s}{2}\bigg[\int_{\R^2}w(\Ri-\Rj) f(\rj,\Rj,\t)d\rj d\Rj\bigg] \deriri \bigg[ f(\ri,\Ri,\t)\bigg]\bigg\}d\ri d\Ri.\label{integral-FP-model 2}
\end{split}
\end{align}

\section*{Acknowledgments}
\noindent The authors thank Martin Burger for the useful discussion
during the Warwick EPSRC symposium workshop on `Emerging PDE models in
socio-eoconomic sciences'. 
The authors are grateful to two anonymous referees for constructive comments and remarks.

\subsection*{Funding} BD has been supported by the Leverhulme Trust research project grant
`Novel discretisations for higher-order nonlinear PDE' (RPG-2015-69).
Part of this research was carried out during a three-month visit of
the second author to the University of Sussex, enabled through financial support by the University of Pavia. MTW acknowledges partial support from the Austrian Academy of Sciences
via the New Frontier's grant NST 0001 and the EPSRC by the grant EP/P01240X/1.

\subsection*{Conflict of Interest} The authors declare that they have no conflict of interest.

\end{document}